\documentclass[preprint,tightenlines,aps,prd,showpacs,nofootinbib]{revtex4}
\usepackage{graphicx}
\usepackage{amsmath}
\usepackage{amssymb}
\usepackage{bm}

\begin{document}

\title{\mbox{}\\[10pt]
Decays of the $X(3872)$ into $J/\psi$ and Light Hadrons
}
%%%%%%%%%%%%%%%%%%%%%%%%%%%%%%%%%%%%%%%%%%%%%%%%%%%%%%%%%%%%%%%%%%%%%%%%%%%%%%

\author{Eric Braaten and Masaoki Kusunoki}
%\thanks{}
\affiliation{
Physics Department, Ohio State University, 
Columbus, Ohio 43210, USA}

\date{\today}
%%%%%%%%%%%%%%%%%%%%%%%%%%%%%%%%%%%%%%%%%%%%%%%%%%%%%%%%%%%%%%%%%%%%%%%%%%%%%%
\begin{abstract}
If the $X(3872)$ is a loosely-bound molecule 
of the charm mesons $D^0 \bar D^{*0}$ and $D^{*0} \bar D^0$,
it can decay through the decay of a constituent in a 
hadronic channel with a nearby threshold, 
such as $J/\psi \, \omega$ or $J/\psi \, \rho$.
The differential decay rates of the $X$ into $J/\psi \, \pi^+\pi^-$,
$J/\psi \, \pi^+\pi^-\pi^0$, $J/\psi \, \pi^0 \gamma$, and
$J/\psi \, \gamma$ are calculated in terms of 
$X J/\psi \, \rho$ and $X J/\psi \, \omega$ coupling constants
using an effective lagrangian that
reproduces the decay rates of the $\omega$ and the $\rho$.
The dependence of the coupling constants on the binding energy 
and the total width of the $X$ is determined by a factorization formula. 
Results from a model by Swanson are used to predict
the partial width of $X$ into $J/\psi \, \pi^+\pi^-\pi^0$
as a function of the binding energy and the total width of the $X$.
\end{abstract}

% 12.38.-t   Quantum chromodynamics
% 12.39.St  Factorization
% 13.20.Gd  Decays of J/psi, Upsilon, and other quarkonia
% 14.40.Gx   Mesons with S=C=B=0, mass > 2.5 GeV (including quarkonia)

%\keywords{}

\maketitle


%%
\section{Introduction}

The $X(3872)$ is a narrow resonance near 
$3872$ MeV discovered by the Belle 
collaboration in electron-positron collisions
through the $B$-meson decay $B^\pm\to XK^\pm$ followed by
the decay $X\to J/\psi\,\pi^+\pi^-$ \cite{Choi:2003ue}. 
Its existence has been confirmed by the CDF and D\O\ collaborations 
through its inclusive production in proton-antiproton 
collisions \cite{Acosta:2003zx,Abazov:2004kp}
and by the Babar collaboration through
the discovery mode $B^\pm\to XK^\pm$ \cite{Aubert:2004ns}. 
The combined measurement of the mass of the $X$ 
is \cite{Olsen:2004fp}
%-----------------
\begin{eqnarray}
 m_X = 3871.9\pm 0.5 \; {\rm MeV}, 
\label{Xmass} 
\end{eqnarray}
%-----------------
which is within 1 MeV of the threshold for the charm mesons 
$D^{0}$  and $\bar D^{*0}$. 
The presence of the $J/\psi$ among the decay products of the $X(3872)$ 
motivates its interpretation as a charmonium state with constituents 
$c\bar c$ \cite{Barnes:2003vb,Eichten:2004uh,Quigg:2004nv,Olsen:2004fp,Quigg:2004vf}.
Two possibilities motivated by the proximity of the mass 
in Eq.~(\ref{Xmass}) to the $D^{0}\bar D^{*0}$ threshold 
are a hadronic molecule with constituents $D D^*$
\cite{Tornqvist,Voloshin:2003nt,Wong:2003xk,Braaten:2003he,Swanson:2003tb,%
Braaten:2004fk,Swanson:2004pp,Voloshin:2004mh,Braaten:2004ai,Braaten:2005jj,%
AlFiky:2005jd}
and a ``cusp'' at the $D^{0}\bar D^{*0}$ threshold associated with 
strong coupling to $D^{0}\bar D^{*0}$ or $D^{*0}\bar D^{0}$
\cite{Bugg:2004rk,Bugg:2004sh}.
Other proposed interpretations include 
a tetraquark with constituents $c \bar c q \bar q$ \cite{Vijande:2004vt},
a ``hybrid charmonium'' state with constituents $c \bar c g$
\cite{Close:2003mb,Li:2004st}, 
a glueball with constituents $ggg$ \cite{Seth:2004zb},
and a diquark-antidiquark bound state with constituents $c u + \bar c \bar u$ 
\cite{Maiani:2004vq}.
The interpretation as a $D D^*$ molecule is particularly predictive
because the small binding energy implies that the molecule has universal
properties that are completely determined by the binding energy
\cite{Voloshin:2003nt,Braaten:2003he,Braaten:2004fk,Braaten:2004ai}.
The small binding energy can be further exploited through factorization 
formulas for production and decay rates of the $X$ \cite{Braaten:2005jj}.

Measurements of the decays of the $X$ can be used to determine 
its quantum numbers and narrow down the possibilities
\cite{Olsen:2004fp,Close:2003sg,Pakvasa:2003ea,Rosner:2004ac,Kim:2004cz,Abe:2005iy}.
The upper bound on the decay width 
of the $X$ is \cite{Choi:2003ue}  
%-----------------
\begin{eqnarray}
 \Gamma_X < 2.3 \;{\rm MeV} \;\;\; {\rm ( 90 \% \; C.L. )},\label{widthX} 
\end{eqnarray}
%-----------------
which is much narrower 
than other charmonium states above 
the $D \bar D$ threshold. 
The product of the branching fractions associated with 
the discovery channel is \cite{Choi:2003ue,Aubert:2004ns,Olsen:2004fp}
%-----------------
\begin{eqnarray}
 {\rm Br}[B^+\to X K^+]\, {\rm Br}[X\to J/\psi\pi^+\pi^-]
 = (1.3 \pm 0.3) \times 10^{-5}.
\label{belle-1}
\end{eqnarray}
%-----------------
The invariant mass distribution of the two pions from the decay
$X \rightarrow J/\psi \, \pi^+\pi^-$ seems to peak near the upper endpoint,
which suggests that the pions come from a virtual $\rho$ resonance. 
Recently the Belle collaboration has observed the $X(3872)$ in the
decay mode $X \rightarrow J/\psi \, \pi^+\pi^-\pi^0$ \cite{Abe:2005iy}
with the branching ratio
%-----------------
\begin{eqnarray}
{ {\rm Br}[X\to J/\psi \, \pi^+\pi^-\pi^0] \over
	{\rm Br}[X\to J/\psi \, \pi^+\pi^-] }
 &=& 1.0 \pm 0.4_{\rm stat} \pm 0.3_{\rm syst}.
\label{BR:pipipi/pipi}
\end{eqnarray} 
%-----------------
The invariant mass distribution of the three pions indicates 
that they come predominantly from a virtual $\omega$ resonance
\cite{Abe:2005ix}.  If the decays into $J/\psi \, \pi^+\pi^-$ 
and $J/\psi \, \pi^+\pi^-\pi^0$ are interpreted as 
$J/\psi \, \rho^{*}$ and $J/\psi \, \omega^*$, 
the approximate equality of the branching fractions in 
Eq.~(\ref{BR:pipipi/pipi}) implies a large violation of isospin symmetry.
The Belle collaboration also reported evidence for 
the decay $X\to J/\psi \, \gamma$ \cite{Abe:2005iy} with the branching ratio
%-----------------
\begin{eqnarray}
{ {\rm Br}[X\to J/\psi \, \gamma] \over
	{\rm Br}[X\to J/\psi \, \pi^+\pi^-] }
 &=& 0.14 \pm 0.05.
\label{BR:gamma/pipi}
\end{eqnarray} 
%-----------------
The observation of the decay into $J/\psi \,\gamma$ 
establishes the charge conjugation of the $X$ to be $+$.
By analyzing angular distributions in the decay of $X$ into 
$J/\psi \, \pi^+ \pi^-$, the Belle collaboration has ruled out
all $J^{P+}$ assignments for $X$ with $J \le 2$ other than 
$1^{++}$ and $2^{++}$ \cite{Abe:2005iy}.
Upper limits have been placed on the branching fractions for
other decay modes of the $X$, including
$D^0 \bar D^0$, $D^+ D^-$, $D^0 \bar D^0 \pi^0$ \cite{Abe:2003zv},
$\chi_{c1} \gamma$, $\chi_{c2} \gamma$, 
$J/\psi \, \pi^0 \pi^0$ \cite{Abe:2004sd},
and $J/\psi \, \eta$ \cite{Aubert:2004fc}.
Upper limits have also been placed on the partial widths for the
decay of $X$ into $e^+e^-$ \cite{Yuan:2003yz,Metreveli:2004px}
and into $\gamma \gamma$ \cite{Metreveli:2004px}. 

The possibility that charm mesons might form molecular states
was considered shortly after the discovery of charm 
\cite{Bander:1975fb,Voloshin:ap,DeRujula:1976qd,Nussinov:1976fg}.  
In 1993, Tornqvist made a quantitative study of the possibility of
molecular states of charm mesons using a 
one-pion-exchange potential model \cite{Tornqvist:1993ng}.
He found that the isospin-0 combinations of $D \bar D^*$ and $D^* \bar D$ 
could form weakly-bound states in the S-wave $1^{++}$ channel 
and in the P-wave $0^{-+}$ channel. 
After the discovery of the $X(3872)$, Tornqvist pointed out that
because the binding energy is small compared to the splitting between  
the $D^+ D^{*-}$ and $D^{0}\bar D^{*0}$ thresholds, there will be
large violations of isospin symmetry \cite{Tornqvist}. 
Swanson considered a potential 
model that includes both one-pion-exchange and quark exchange
and found that the $C=+$ superposition of $D^{0}  \bar D^{*0}$
and $D^{*0} \bar D^{0}$ could form a weakly-bound state 
in the S-wave $1^{++}$ channel \cite{Swanson:2003tb}.
Swanson's model included not only the charmed mesons
$D^{0}  \bar D^{*0}$, $D^{*0} \bar D^{0}$, $D^+ D^{*-}$,
and $D^{*+} D^{-}$, but also two other pairs of hadrons
with nearby thresholds: $J/\psi \, \rho^0$ and $J/\psi \, \omega$.
His prediction that the branching fraction for the decay of 
$X$ into $J/\psi \, \pi^+ \pi^- \pi^0$ should be comparable to that 
for decay into $J/\psi \, \pi^+ \pi^-$ was verified 
by the Belle collaboration \cite{Abe:2005ix,Abe:2005iy}.

In this paper, we analyze decays of the $X(3872)$ into $J/\psi$ 
and light hadrons under the assumption that $X$ is a loosely-bound 
$D D^*$ molecule and that these decays proceed through transitions 
of $X$ to $J/\psi \, \rho$ and $J/\psi \, \omega$.
In Section~\ref{sec:DDstar}, we summarize some of the universal results 
for a system with large scattering length and we give the current 
constraints on the real and imaginary parts of the large scattering 
length for the $D D^*$ system.
In Section~\ref{sec:nearby}, we discuss other hadronic states with 
thresholds near the $D^{0}\bar D^{*0}$ threshold and we summarize 
results from Swanson's model for the $X(3872)$.
In Section~\ref{sec:Xdecay}, we calculate the differential
distributions for decays of $X$ into
$ J/\psi \, \pi^+\pi^-$, $J/\psi \, \pi^+\pi^-\pi^0$, and
$J/\psi \, \pi^0 \gamma$ using an effective lagrangian that reproduces 
decays of the vector mesons.  We also use vector meson dominance to 
calculate the partial width for the decay into $J/\psi \, \gamma$.
The normalizations of the decay rates are determined by unknown 
coupling constants for the transitions of $X$ to $J/\psi \, \rho$ 
and $J/\psi \, \omega$.  The dependence of the coupling constants 
on the binding energy and the total width of the $X$ 
is deduced using a factorization formula.
In Section~\ref{sec:model}, we use a two-channel scattering
model to illustrate how the coupling constants can be related to
probabilities for the $J/\psi \, \rho$ and $J/\psi \, \omega$
components of the $X$.
We use the probability for $J/\psi \, \omega$
in Swanson's model to give a quantitative 
prediction for the partial decay rate of the $X$ into 
$J/\psi \, \pi^+\pi^-\pi^0$ as a function of the binding energy 
and the total width of the $X$.
A summary of our results is given in Section~\ref{sec:summary}.
An updated determination of the parameters in the effective lagrangian 
for light pseudoscalar and vector mesons is given in an Appendix. 

%\newpage

\section{Universality and the $D D^*$ System} 
\label{sec:DDstar}

The mass of the $X$ is extremely close to the 
$ D^{0} \bar D^{*0}$ threshold: 
$m_{D^{0}}+m_{D^{*0}}=3871.3\pm 1.0$ MeV. 
From the mass measurement in Eq.~(\ref{Xmass}), 
the difference is
%-----------------
\begin{equation}
m_{X} - (m_{D^{0}}+m_{D^{*0}})= +0.6\pm 1.1  \; {\rm MeV}.
\label{mXdiff}
\end{equation}
%-----------------
Most of the uncertainty comes from the experimental uncertainty in 
$2 m_{D^0}$, because the mass difference $m_{D^{*0}}-m_{D^{0}}$
has a much smaller uncertainty.
If the $X$ were a $D^{0}\bar D^{*0}$/$D^{*0}\bar D^{0}$ molecule, 
the energy difference in Eq.~(\ref{mXdiff}) would have to be 
negative, corresponding to a positive binding energy defined by 
%-----------------
\begin{equation}
E_X = (m_{D^{0}} + m_{D^{*0}}) - m_{X}.
\label{EX}
\end{equation}
%-----------------
The measurement of the binding energy in Eq.~(\ref{mXdiff}) is 
compatible with a small negative value 
corresponding to a $D D^*$ molecule. 
However, the central value of the energy difference
in Eq.~(\ref{mXdiff}) is positive, corresponding 
to a resonance in $D^0\bar D^{*0}$ and $D^{*0}\bar D^{0}$ scattering
rather than a bound state. 
Bugg has referred to this possibility as a ``cusp state'' 
\cite{Bugg:2004rk,Bugg:2004sh}, because the line shape of the $X$ 
in some of its decay modes has a cusp at the $D^{*0}\bar D^{0}$ threshold.

The energy difference in Eq.~(\ref{mXdiff})
is tiny compared to the natural energy scale for binding by
the pion exchange interaction: $m_{\pi}^2/2\mu \approx 10$ MeV, 
where $\mu$ is the reduced mass of $D^0$ and $\bar D^{*0}$:
%-----------------
\begin{eqnarray}
 \mu = \frac{ m_{D^{0}} m_{D^{*0}} }
{m_{D^{0}}+m_{D^{*0}}} = 966.5 \pm 0.3 \; {\rm MeV} .
\label{reducedmass}
\end{eqnarray} 
%-----------------
Whether the energy difference is positive or negative,
its unnaturally small value implies that if the $X$ couples to 
$ D^{0} \bar D^{*0}$ and $D^{*0} \bar D^{0}$,
the S-wave scattering lengths for those channels
must be large compared to the natural length scale $1/m_{\pi}$
associated with the pion exchange interaction.
Since the experimental evidence favors the charge conjugation 
quantum number $C=+$, we assume that there is a large scattering 
length $a$ in the $C=+$ channel and that the scattering length 
in the $C=-$ channel is negligible in comparison. 
In this case, the scattering lengths for elastic 
$D^{0} \bar D^{*0}$ scattering and elastic 
$D^{*0} \bar D^{0}$ scattering are both $a/2$.

Nonrelativistic few-body systems with short-range interactions
and a large scattering length have universal properties that
depend on the scattering length but are otherwise insensitive  
to details at distances small compared to $|a|$ \cite{Braaten:2004rn}. 
We consider the scattering length to be large if it is much larger
than the natural momentum scale associated with low-energy scattering.
The universal results are encoded in the truncated connected 
transition amplitude, which is a function of the total energy 
$E$ of the two particles in the rest frame of the pair:
%-----------------
\begin{equation}
{\cal A}(E) = {2 \pi/\mu \over - 1/a + \sqrt{-2 \mu E} },
\label{A-uni}
\end{equation}
%-----------------
where $\mu$ is the reduced mass of the two particles.
If $a$ is real and positive, this amplitude has a pole on the 
physical sheet at $E = -1/(2 \mu a^2)$, indicating the existence 
of a weakly-bound state with the universal binding energy  
%-----------------
\begin{equation}
E_X= \frac{1}{2\mu a^2}.
\label{Eb}
\end{equation}
%-----------------
The universal momentum-space wavefunction of this bound state is
%-----------------
\begin{eqnarray}
\psi(p) = \frac{(8\pi/a)^{1/2}}{p^2+1/a^2}.
\label{psi-uni}
\end{eqnarray}
%-----------------
The universal amplitude for transitions from the bound state
to a scattering state consisting of two particles with small 
relative momentum is determined by the residue of the pole 
in ${\cal A}(E)$:
%-----------------
\begin{equation}
{\cal A}_X = {\sqrt{2 \pi} \over \mu} \, a^{-1/2}.
\label{AX-uni}
\end{equation}
%-----------------

If there is an inelastic scattering channel,
the large scattering length $a$ has a negative imaginary part.
It is convenient to express the complex scattering length
in the form
%-----------------
\begin{eqnarray}
 \frac{1}{a}= \gamma_{\rm re} + i \gamma_{\rm im},
\label{a-complex}
\end{eqnarray} 
%-----------------
where $\gamma_{\rm re}$ and $\gamma_{\rm im}$ are real 
and $\gamma_{\rm im}\ge 0$.
The universal expression for the binding energy in Eq.~(\ref{Eb})
has an imaginary part $i \Gamma_X/2$, where 
%-----------------
\begin{eqnarray}
\Gamma_X  = 
2\gamma_{\rm re} \gamma_{\rm im}/\mu .
\label{Gam-bs}
\end{eqnarray}
%-----------------
If $\gamma_{\rm im} < \gamma_{\rm re}$, $\Gamma_X$ is the full width 
at half maximum of a resonance in the inelastic channel \cite{Braaten:2005jj}.
It therefore can be interpreted as the rate for the 
decay of the bound state into the inelastic channel.
The peak of the resonance is below 
the threshold by the amount
%-----------------
\begin{eqnarray}
E_X = \gamma_{\rm re}^2/(2 \mu) . 
\label{E-bs}
\end{eqnarray}
%-----------------
We can therefore interpret this expression as the binding energy 
of the resonance \cite{Braaten:2005jj}.

The observed decays of the $X$ imply that there are inelastic 
scattering channels, so $a$ has a negative imaginary part.
It can be parameterized in terms of the 
real and imaginary parts of $1/a$ as in Eq.~(\ref{a-complex}). 
Our interpretation of $X$ as a bound state requires $\gamma_{\rm re} > 0$.
The energy difference in Eq.~(\ref{mXdiff}) puts an upper bound on
$\gamma_{\rm re}$:
%-----------------
\begin{eqnarray}
\gamma_{\rm re} < 40 \ {\rm MeV} \hspace{1cm} ({\rm 90\% \ C.L.}).
\label{gamre:ub}
\end{eqnarray}
%-----------------
The upper bound on the width in Eq.~(\ref{widthX}) 
puts an upper bound on the product of $\gamma_{\rm re}$ and 
$\gamma_{\rm im}$:
%-----------------
\begin{eqnarray}
\gamma_{\rm re} \gamma_{\rm im} < (33 \ {\rm MeV})^2 
\hspace{1cm} ({\rm 90\% \ C.L.}).
\end{eqnarray}
%-----------------
There is also a lower bound on the width of the $X$
from its decays into $D^0 \bar D^0 \pi^0$ and $D^0 \bar D^0 \gamma$, 
which both proceed through the decay of a constituent $D^*$.
These decays involve interesting interference effects,
but the decay rates have smooth limits as the binding energy 
is tuned to 0 \cite{Voloshin:2003nt}.  In this limit, the constructive
interference increases the decay rate by a factor of 2, so
the partial widths of $X$ into $D^0 \bar D^0 \pi^0$ 
and $D^0 \bar D^0 \gamma$ add up to $2 \, \Gamma[D^{*0}]$.  
The width of $D^{*0}$ has not been measured, 
but it can be deduced from other information 
about the decays of $D^{*0}$ and $D^{*+}$.
Using the total width of the $D^{*+}$, 
its branching fraction into $D^+ \pi^0$, and isospin symmetry,
we can deduce the partial width of $D^{*0}$ into $D^0 \pi^0$ to be
$42 \pm 10$ keV.
The total width of the $D^{*0}$ can then be obtained by 
dividing by its branching fraction into $D^0 \pi^0$:
$\Gamma[D^{*0}] = 68 \pm 16$ keV.
The sum of the partial widths of $X$ into $D^0 \bar D^0 \pi^0$ and 
$D^0 \bar D^0 \gamma$ is therefore $136 \pm 32$ keV.
The resulting lower bound on the product 
of $\gamma_{\rm re}$ and $\gamma_{\rm im}$ is 
%-----------------
\begin{eqnarray}
\gamma_{\rm re} \gamma_{\rm im} > (7 \ {\rm MeV} )^2
\hspace{1cm} ({\rm 90\% \ C.L.}).
\end{eqnarray}
%-----------------
By combining this with the upper bound on $\gamma_{\rm re}$
in Eq.~(\ref{gamre:ub}), we can infer that $\gamma_{\rm im}> 1$ MeV.

%\newpage

\section{Hadronic states with nearby thresholds}
\label{sec:nearby}

The state of the $X$ can be written schematically as
%-----------------
\begin{eqnarray}
 |X\rangle = {Z^{1/2} \over \sqrt{2}}
\left( | D^{0} \bar D^{*0} \rangle 
+ |D^{*0} \bar D^{0} \rangle \right)
+ \sum_{H} Z^{1/2}_{H} |H\rangle,   
\label{state-X}
\end{eqnarray}
%-----------------
where $Z$ is the probability for the $X$ to be in 
the $ D^{0} \bar D^{*0}/D^{*0} \bar D^{0}$ state
and $Z_{H}$ is the probability for the $X$ to be in 
another hadronic state $H$.
The hadronic states $H$ 
in (\ref{state-X}) could include charmonium states, 
other charm meson pairs such as $D^\pm D^{*\mp}$, 
states consisting of a charmonium and a light hadron 
such as $J/\psi \,  \rho^{}$ and $J/\psi \, \omega$, etc. 
An expansion of $|X\rangle$ in terms of hadronic states 
can be valid only if there is an ultraviolet cutoff 
on the energy difference with respect to the $D^0 \bar D^{*0}$ 
threshold.  The probabilities $Z_H$ depend on that cutoff.
Universality implies that as 
$a\rightarrow \infty$, $Z$ approaches to $1$ and 
$Z_{H}$ scales as $1/a$  \cite{Braaten:2003he}. 

The small binding energy of $X$ compared to the natural energy scale
$m_\pi^2/(2\mu) = 10$ MeV associated with pion exchange implies 
resonant S-wave interactions in the $ D^{0} \bar D^{*0}$ and 
$D^{*0} \bar D^{0}$ systems.
If there are other hadronic channels whose thresholds differ from the 
$D^0 \bar D^{*0}$ threshold by less than 10 MeV, there would be
resonant S-wave interactions in those channels as well.
In this case, it would be necessary to treat 
all the resonating channels as a coupled-channel system,
with a large elastic scattering length for each channel and a large 
transition scattering length  for each pair of channels. 
The hadronic states $D^\pm D^{*\mp}$, $J/\psi\, \rho$, and 
$J/\psi\,\omega$ have thresholds that are relatively close to the 
$D^{0}\bar D^{*0}$ threshold. 
The energy gaps between these other thresholds 
and the $D^{0} \bar D^{*0}$ threshold are 
%-----------------
\begin{subequations}
\begin{eqnarray}
m_{D^{\pm}} + m_{D^{*\mp}}- (m_{D^0}+m_{D^{*0}})
 &=& +8.1 \pm 0.1  \;{\rm MeV},  
\label{gap-DDstar}  
\\
m_{J/\psi} + m_{\rho} - ( m_{D^0}+m_{D^{*0}})
 &=& +1.4 \pm 1.1  \;{\rm MeV},  
\label{gap-rho}  
\\
m_{J/\psi} + m_{\omega} - ( m_{D^0}+m_{D^{*0}})
 &=& +8.2 \pm 1.0  \;{\rm MeV}. 
\label{gap-omega}
\end{eqnarray}
\label{gap-eq}
\end{subequations}
%-----------------
The small uncertainty in Eq.~(\ref{gap-DDstar}) comes from using mass
differences between charm mesons to calculate the energy gap.
The uncertainties in Eqs.~(\ref{gap-rho}) and (\ref{gap-omega})
are dominated by the uncertainty in $2 m_{D^0}$.
The energy gaps in the $D^\pm D^{*\mp}$ and $J/\psi\,\omega$ channels 
are comparable to the natural energy scale of about $10$ MeV 
associated with pion exchange.
The energy gap in Eq.~(\ref{gap-rho}) for the $J/\psi\, \rho$ channel
is much smaller.  However
whether any of these channels can have resonant interactions 
with $D^{0}\bar D^{*0}$ or $D^{*0}\bar D^{0}$ is determined 
not only by the real parts of the energy gaps, which are given in 
Eqs.~(\ref{gap-eq}), but also by the imaginary parts, which can be
obtained by replacing each mass $m$ by $m-i\Gamma/2$, 
where $\Gamma$ is the width of the particle. 
If there are large differences between the widths of the various 
particles, it is necessary only to take into account the largest 
width among the particles in each channel. 
The largest width in each of the three channels is
%-----------------
\begin{subequations}
\begin{eqnarray}
 \Gamma [ D^{*\pm} ]   &=& 0.096 \pm 0.022 \; {\rm MeV}, 
 \\
 \Gamma [\rho    ]     &=& 150.3 \pm 1.6 \; {\rm MeV}, 
 \\
 \Gamma [\omega ]      &=&  8.49 \pm 0.08 \; {\rm MeV}.
\label{gam-omega}
\end{eqnarray}
\end{subequations}
%-----------------
For the $D^\pm D^{*\mp}$ and $J/\psi \, \omega$ channels,
the magnitude $|\Delta|$ of the complex energy gap is comparable to 
the natural energy scale 10 MeV associated with pion
exchange between $D$ and $D^*$.
The large width of the $\rho$ makes $|\Delta|$ for the $J/\psi \, \rho$ 
channel much larger than the natural energy scale.  

Because the complex energy gap $\Delta$ for the other hadronic channels 
with nearby thresholds are comparable to or larger than the natural 
energy scale, these channels need not be taken into account explicitly 
in calculations of quantities that have nontrivial universal limits 
as $a \to \pm \infty$.  Their dominant effects enter through the 
complex-valued scattering length $a$.
A coupled-channel model that includes other hadronic states with 
nearby thresholds could still be useful for estimating nonuniversal 
quantities or for calculating nonuniversal corrections to the 
universal predictions.  

Swanson has constructed a model of the $X$ and the hadronic states 
with nearby thresholds and used it to predict some of the properties 
of the $X$ \cite{Swanson:2003tb,Swanson:2004pp}.
In particular, he predicted correctly that the branching fraction 
for $X \to J/\psi \, \pi^+ \pi^- \pi^0$ is comparable to that for 
$X \to J/\psi \, \pi^+ \pi^-$.
In addition to the channel $D^{0} \bar D^{*0}+ D^{*0} \bar D^{0}$,  
Swanson's model includes $D^{+}  D^{*-}+D^{*+}  D^{-}$,
$J/\psi\, \rho$, and $J/\psi\, \omega$. 
It includes the S-wave and D-wave channels for $D D^*$, 
but only the S-wave channel for $J/\psi \, V$,
where $V$ is the vector meson $\rho$ or $\omega$.
Thus the model has 6 coupled channels.
The interactions between the hadrons are modeled by potentials:
one-pion-exchange potentials for the S-wave and D-wave $D D^*$
channels and for transitions between those channels
and Gaussian potentials for the transitions between 
the S-wave $D D^*$ channels and the S-wave $J/\psi \, V$ channels
to simulate the effects of quark exchange.
The one-pion-exchange potential is singular at short distances 
and it was regularized by an ultraviolet momentum cutoff $\Lambda$.
The nonrelativistic Schr\"odinger equation for the 6 coupled channels
was solved numerically. 
A bound state with the quantum numbers $J^{PC}=1^{++}$ of $X$
appeared when the ultraviolet cutoff exceeded the critical value
$\Lambda_c = 1.45$ GeV. 
The binding energy of the $X$ could be adjusted by varying the 
ultraviolet cutoff.

Swanson solved the coupled channel problem under the assumption that
the $\rho$ and $\omega$ are 
stable hadrons with equal masses $m_\rho=m_\omega=782.6$ MeV.
The reason for using an unphysical value for $m_\rho$ 
is that the central PDG value from 2002 and earlier,
$m_\rho = 771.1$ MeV, is below the $D^0 \bar D^{*0}$ threshold.
If such a value had been used, it would have been necessary to 
treat $J/\psi \, \rho$ states as scattering states.
This complication was avoided by using a value of $m_\rho$
above the $D^0 \bar D^{*0}$ threshold.
Note that in Eq.~(\ref{gap-rho}), we have taken the updated 
2004 PDG value $m_{\rho}=775.8\pm 0.5$ MeV \cite{Eidelman:2004wy}, 
which gives a $J/\psi\, \rho$ threshold that is a few MeV 
above the mass of the $X$. 

%------------------------------------------------------------------------------------
\begin{figure}[t]
\includegraphics[width=7cm]{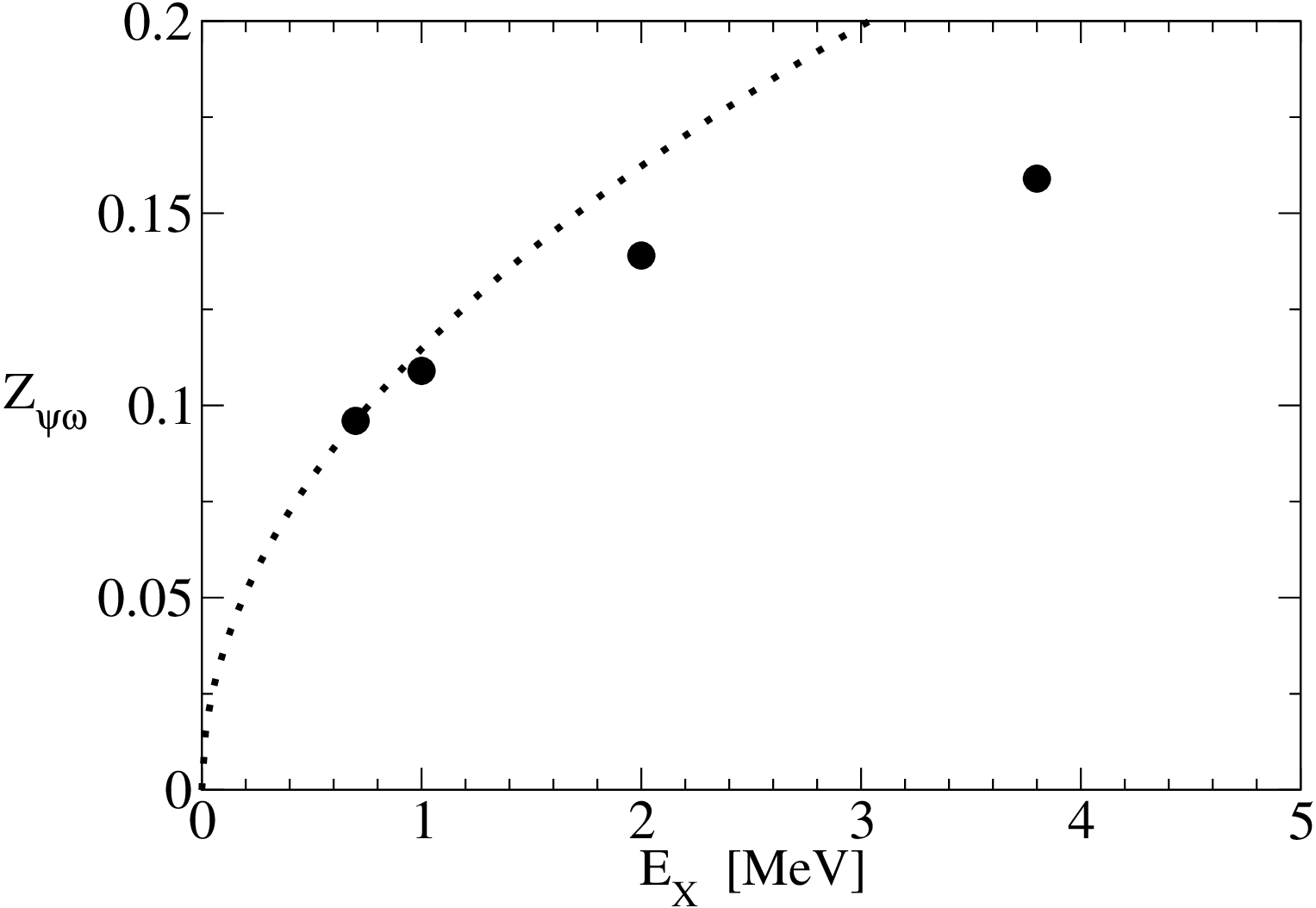}
\hspace{0.5cm}
\includegraphics[width=7cm]{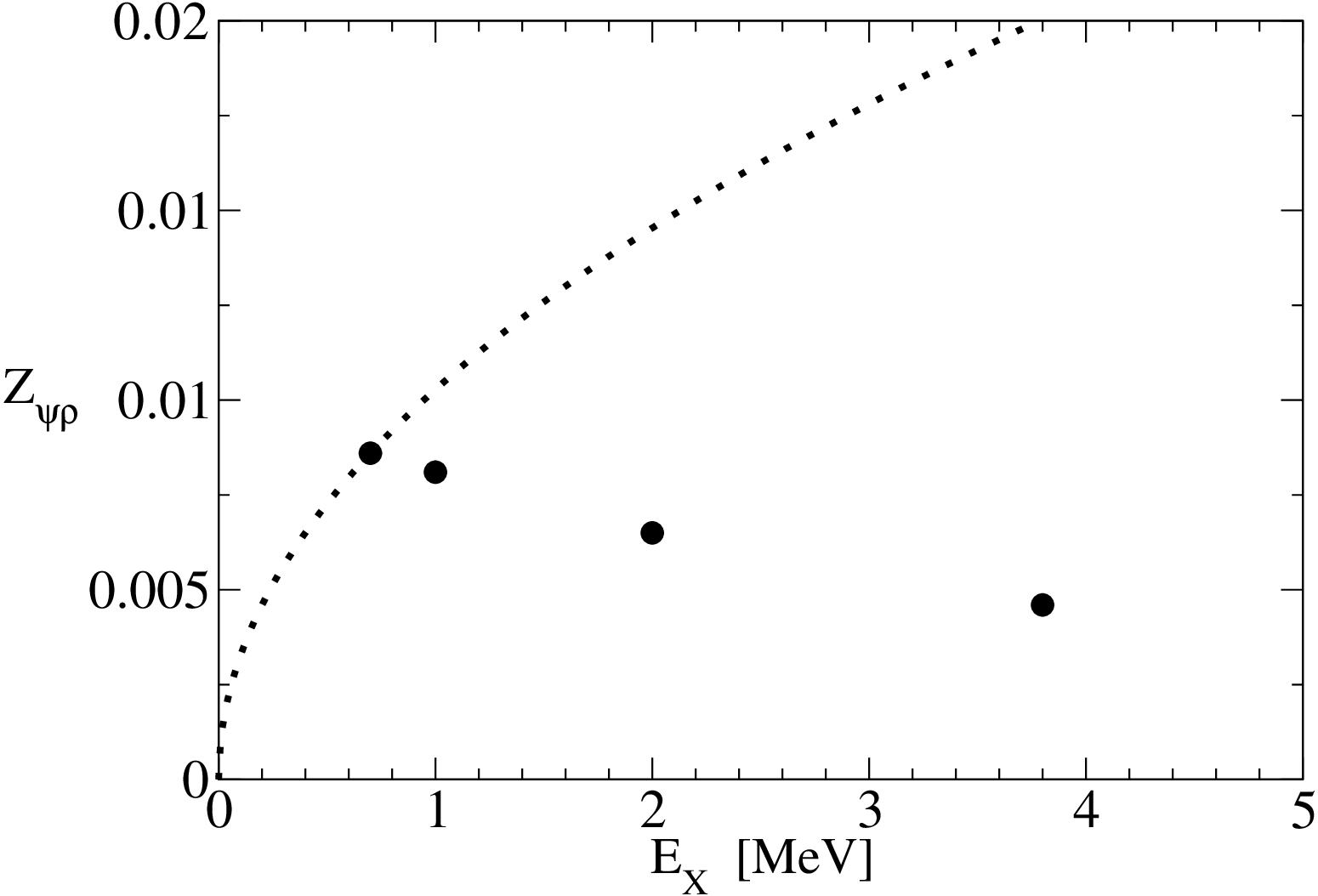}
\caption{The probabilities $Z_{\psi\omega}$ and $Z_{\psi\rho}$
for the $J/\psi\,\omega$ and $J/\psi\,\rho$ components of the $X(3872)$ 
as a function of its binding energy $E_X$. 
The dots are the results from Swanson's model \cite{Swanson:2003tb}. 
The dotted curves have the scaling behavior $E_X^{1/2}$ 
and pass through the model result for $E_X = 0.7$ MeV.
\label{fig:matching}}
\end{figure}
%------------------------------------------------------------------------------------

Swanson calculated the probabilities for 
each component of the wavefunction of $X$ for values of 
the ultraviolet cutoff that correspond to varying the binding
energy $E_X$ from 0.7 MeV to 23.2 MeV \cite{Swanson:2003tb}.
His results for the probabilities $Z_{\psi\omega}$ and $Z_{\psi\rho}$
are shown as dots in Fig.~\ref{fig:matching}. 
Since the binding energy of the $X$ is known to be less than 1 MeV, 
only the lowest two values of $E_X$ could be physically relevant.
For the lowest value $E_X= 0.7$ MeV, the probabilities were 
$Z_{\psi \omega} = 9.6$\%, $Z_{D^\pm D^{*\mp}} = 7.9$\%, 
and $Z_{\psi \rho} = 0.86$\%.  The total probability for the 
$D^0 \bar D^{*0}$ and $D^{*0}\bar D^0$ components of the
wavefunction is 81.6\%.

In Fig.~\ref{fig:matching}, the dotted lines have the scaling behavior 
$E_X^{1/2}$ predicted by universality and are normalized so that
they pass through the dot at $E_X = 0.7$ MeV.  
The probability $Z_{\psi \omega}$ clearly exhibits the universal behavior.
The probability $Z_{\psi \rho}$ does not.  This may be related 
to the fact that $Z_{\psi \rho}$ is more than an order of magnitude 
smaller than $Z_{\psi \omega}$.  Because of the weaker coupling 
of the $X$  to isospin-1 states, the scaling region for isospin-1 
states may not set in until a much smaller value of $E_X$.

Swanson estimated the partial widths for the decays of $X$ into  
$J/\psi\, h$, where $h$ is the light hadronic state $\pi^+\pi^-$, 
$\pi^+\pi^-\pi^0$, $\pi^0\gamma$, or $\pi^+\pi^-\gamma$,
using a simple ad hoc recipe.
The partial width into  $J/\psi\, h$ was taken to be the sum over 
the vector mesons $V = \rho,\omega$ of the product of the 
probability $Z_{\psi V}$ for the $J/\psi\, V$ component of the 
wavefunction and the partial width $\Gamma[V \to h]$ for the decay 
of the vector meson:
%-----------------
\begin{equation}
\Gamma[X \to J/\psi \, h] \approx
\sum_V Z_{\psi V} \Gamma[V \to h].
\label{gam-Swanson}
\end{equation}
%-----------------
For the smallest value of the binding energy that was considered,
$E_X= 0.7$ MeV, the resulting estimates of the partial widths 
for decay into $J/\psi\, h$ were 1290 keV, 720 keV, 70 keV, 
and 13 keV for $h=\pi^+\pi^-$, 
$\pi^+\pi^-\pi^0$, $\pi^0\gamma$, and $\pi^+\pi^-\gamma$, respectively.
The ratio of the partial widths into $J/\psi\, \pi^+\pi^-\pi^0$
and $J/\psi\, \pi^+\pi^-$ was predicted to be 0.56 for $E_X= 0.7$ MeV.  
Remarkably, this prediction agrees with the subsequent measurement 
by the Belle collaboration given in Eq.~(\ref{BR:pipipi/pipi}) 
to within the experimental errors \cite{Abe:2004sd}.  
The approximately equal branching fractions are a
fortuitous result of an amplitude for $X \to J/\psi \, \omega$
that is much larger than the amplitude for $X \to J/\psi \, \rho$
and an amplitude for $\omega \to \pi^+ \pi^- \pi^0$ that 
is much smaller than that for $\rho \to \pi^+ \pi^-$.
The suppression of the amplitude for $X \to J/\psi \, \rho$
is related to the fact that in the isospin symmetric limit in
which the mass difference between neutral and charged
$D$'s is neglected, there is binding in the isospin-0 channel 
but not in the isospin-1 channel \cite{Tornqvist:1993ng}.

Swanson has also used his model to calculate the rates for several 
other decay modes of the $X$ \cite{Swanson:2004pp}.  
The decay rate into $J/\psi \, \gamma$
has contributions from transitions to $J/\psi \, \rho$ 
and $J/\psi \, \omega$ that can be calculated using vector meson 
dominance.  It also has contributions from the annihilation 
of the $u$ and $\bar u$
from the charm mesons that are the constituents of the $X$.
Swanson's prediction for the partial width into $J/\psi \, \gamma$
for an $X$ with a binding energy of 1 MeV is 8 keV.
Decay modes that receive contributions only from $u \bar u$ 
annihilation, such as $\psi(2S) \, \gamma$, $K K^*$, and  $\pi \rho$,
have much smaller partial widths.

%\newpage 

\section{Decays of $\bm{X}$ into $\bm{J/\psi \, h}$} 
\label{sec:Xdecay}

In this section, we calculate the differential decay rates of the $X$ into 
$J/\psi\, h$, where the hadronic system $h$ is $\pi^+ \pi^- \pi^0$,
$\pi^+ \pi^-$, $\pi^0 \gamma$, or $\gamma$.
We assume that these decays proceed through transitions of $X$ to
$J/\psi \, \rho$ and $J/\psi \, \omega$.  We calculate the 
differential decay rates in terms of two unknown complex coupling  
constants using an effective lagrangian that reproduces the decays 
of the light vector mesons.

\subsection{Vector Meson Decay Amplitudes}

We assume that the decay of $X$ into $J/\psi\, h$, where $h$ is a system 
of light hadrons, proceeds through transitions to $J/\psi \, V$,
where $V$ is one of the vector mesons $\rho$ or $\omega$,
followed by the decay of the vector meson into $h$.
Because the mass of the $X$ is so close to the threshold for 
$J/\psi \, V$, the vector meson is almost on its mass shell.
Any model that reproduces the decays of the vector mesons 
should also accurately describe the decay of the virtual vector meson 
in the $J/\psi \, V$ component of $X$.
In Ref.~\cite{Braaten:1989zn},
the semileptonic branching fractions for the $\tau$ lepton were
calculated using an effective lagrangian for light pseudoscalar 
and vector mesons with $U(3) \times U(3)$ chiral symmetry.  
All the parameters in the effective lagrangian,
aside from the pion decay constant, were 
determined directly from decays of the vector mesons $\rho$ and 
$\omega$.   That same effective lagrangian can be used to calculate the 
partial widths of $X$ into $J/\psi\, h$.
An updated determination of the parameters in that
effective lagrangian is given in the Appendix.

The T-matrix element for the decay of a vector meson $V$ into 
the light hadronic state $h$ 
can be expressed in the form
%-----------------
\begin{eqnarray}
 {\cal T}[V \to h] 
&=& 
\epsilon_V^{\mu} 
{\cal A}_{\mu}[ V \to h], 
\end{eqnarray}
%-----------------
where $\epsilon_V$ is the polarization vector of the vector meson. 
The amplitude ${\cal A}_{\mu}$ for the decay 
$\rho \to \pi^+\pi^-$ is    
%-----------------
\begin{eqnarray}
 {\cal A}_{\mu}[\rho\to \pi^+\pi^-] =
  \mbox{$1\over 2$} \, 
G_{v \pi \pi} \,  
 \left(p_{+}-p_{-}\right)_{\mu}.
\label{amp-rho}
\end{eqnarray}
%-----------------
The value of the coupling constant $G_{v \pi \pi}$ 
is given in Eq.~(\ref{Gvpipi}).
The amplitude ${\cal A}_{\mu}$ for the decay 
$\omega\to \pi^+\pi^-\pi^0$ is
%-----------------
\begin{eqnarray}
 {\cal A}_{\mu}[\omega\to \pi^+\pi^-\pi^0] 
&=& 
\frac{4\sqrt{3} (\cos\theta_v+\sqrt{2}\sin\theta_v)}{F_{\pi}^3} \,
\varepsilon_{\mu\nu\alpha\beta} \, p_{+}^{\nu} \, p_{-}^{\alpha} \, p_{0}^{\beta} \,
\nonumber \\
&& \hspace{-2.0cm}
\times
\left(C_{v 3\pi}
+ \frac{ G_{v \pi \pi}C_{vv\pi}F_{\pi}^2 } { m_v^2 } 
\left( 1 -  \mbox{$1\over 3$} \,  
\left[f_\rho(s_{12})+f_\rho(s_{23})+f_\rho(s_{31})\right]
 \right) \right),
\label{amp-omega}
\end{eqnarray}
%-----------------
where $s_{12}$, $s_{23}$, and $s_{31}$ are the invariant masses
of the three different pion pairs and
%-----------------
\begin{equation}
f_{V}(s)\equiv {s \over 
s-M_{V}^2+im_{V}\Gamma_{V}}
\end{equation}
%-----------------
is a vector meson resonance factor that vanishes at $s=0$. 
We have denoted the 4-momenta of $\pi^+$, $\pi^-$, and $\pi^0$
by $p_+$, $p_-$, and $p_0$, respectively.  
The pion decay constant is $F_{\pi}= 93$ MeV,
the values of the parameters $C_{v 3\pi}$ and 
$G_{v \pi \pi}C_{vv\pi}F_{\pi}^2/m_v^2$
are given by Eqs.~(\ref{Cv3pi}) and (\ref{GCvvpi}),
and the value of the light vector meson mixing angle $\theta_{v}$ is 
given by Eq.~(\ref{costheta}).
The amplitudes ${\cal A}_{\mu}$ for the radiative decays 
of the vector mesons are
%-----------------
\begin{subequations}
\begin{eqnarray}
{\cal A}_{\mu}[\rho^+\to \pi^+\gamma]
&=&
 {4 e \over 3 F_\pi}
 \,
 \left( C_{v\pi\gamma}+ {G_{v\gamma}\,C_{vv\pi}\,
 F_\pi^2 \over m_v^2 } \right)
 \, \varepsilon_{\mu\nu\alpha\beta} 
 Q^{\nu} p^{\alpha}\epsilon_{\gamma}^{\beta} ,
\\
{\cal A}_{\mu}[\omega\to \pi^0\gamma]
&=&
 {4 (\cos\theta_v + \sqrt{2} \sin \theta_v) e \over \sqrt{3} F_\pi}
 \,
 \left( C_{v\pi\gamma}+ {G_{v\gamma}\,C_{vv\pi}\,
 F_\pi^2 \over m_v^2 } \right)
 \, \varepsilon_{\mu\nu\alpha\beta} 
 Q^{\nu} p^{\alpha}\epsilon_{\gamma}^{\beta} ,
\label{amp:pigamma-omega}
\end{eqnarray}
\label{amp:pigamma}
\end{subequations}
%-----------------
where $Q$ and $p$ are the 4-momenta of the vector meson 
and the pion and $\epsilon_\gamma$
is the polarization vector of the photon.
The values of the parameters $C_{v\pi\gamma}$ and 
$G_{v\gamma} C_{vv\pi}F_{\pi}^2/m_v^2$
are given by Eqs.~(\ref{Cvpigam}) and (\ref{GCvgam}).

The amplitudes ${\cal A}_\nu$ in Eqs.~(\ref{amp-rho}),
(\ref{amp-omega}), and (\ref{amp:pigamma}) all satisfy
$Q^\nu {\cal A}_\nu=0$, where $Q$ is the 4-momentum 
of the vector meson.  This condition is satisfied in any model
consistent with vector meson dominance.  
The assumption of vector meson dominance is that the amplitude 
for the production of a real photon in a hadronic process 
can be expressed as the sum of over vector mesons $V$ of the 
amplitude for producing $V$ multiplied by a 
coupling constant for the transition $V \to \gamma$.  
The condition $Q^\nu {\cal A}_\nu=0$
is required for the gauge invariance of the resulting 
amplitude for real photon production.

The T-matrix element for $X$ to decay into $J/\psi$ and 
a light hadronic system $h$ 
through a virtual vector meson resonance $V$ can be expressed as
%-----------------
\begin{eqnarray}
{\cal T}[X\to J/\psi \, h] 
= {\cal A}_\mu[X \to J/\psi \, V]
{- g^{\mu \nu} \over Q^2 - m_V^2 + i m_V \Gamma_V}
{\cal A}_\nu[ V \to h],
\label{T-psipi}
\end{eqnarray} 
%-----------------
where $Q$ is the total 4-momentum of the hadronic system $h$
or, equivalently, of the virtual vector meson.
We have used the condition $Q^\nu {\cal A}_\nu=0$
to simplify the numerator of the vector meson propagator.
The quantum numbers of the particles, together with Lorentz 
invariance, constrains the amplitude for $X \to J/\psi \, V$
to be the sum of two terms.  One of them is
%-----------------
\begin{equation}
{\cal A}_{\mu}[X \to J/\psi \, V]
=  G_{X \psi V} \, 
\varepsilon_{\mu \nu \alpha \beta} Q^\nu \epsilon_X^\alpha
\epsilon_{\psi}^{*\beta} ,
\label{amp-jpsiV}
\end{equation}
%-----------------
where $\epsilon_X$ and $\epsilon_{\psi}$ are the
polarization 4-vectors of the $X$ and the $J/\psi$
and $G_{X \psi V}$ is a dimensionless constant.
The contraction of this amplitude with the polarization vector
$\epsilon_V^*$ of the vector meson reduces in the rest frame 
of the vector meson to 
$G_{X \psi V} m_V \bm{\epsilon_X} \cdot 
	(\bm{\epsilon}_{\psi} \times \bm{\epsilon}_V)^*$.
The other independent amplitude ${\cal A}_{\mu}$
has the Lorentz structure	
$\varepsilon_{\mu \nu \alpha \beta} P^\nu \epsilon_X^\alpha
\epsilon_{\psi}^{*\beta}$.  In the rest frame of the $X$, 
its contraction with $\epsilon_V^*$ is 	
$m_X \bm{\epsilon_X} \cdot 
	(\bm{\epsilon}_{\psi} \times \bm{\epsilon}_V)^*$.
Since the mass of the $X$ is so close to the threshold for  
$J/\psi \, V$, the rest frames of the $X$ and $V$ are essentially
identical.  Thus the two independent Lorentz structures 
are essentially equivalent for decays that are dominated 
by the vector meson resonance.  They give similar
predictions for the partial widths for $X$ into $J/\psi \, h$ for 
$h = \pi^+ \pi^- \pi^0$, $\pi^+ \pi^-$, or $\pi^0 \gamma$.
The amplitude in Eq.~(\ref{amp-jpsiV}) has the advantage 
that it is also consistent with the constraint
$Q^\mu {\cal A}_\mu=0$ required by vector meson dominance.  
Thus this amplitude can be used to calculate the  
decay of $X$ into $J/\psi \, \gamma$.  We therefore take the 
transition amplitude for $X$ into $J/\psi \, V$
to be the expression in Eq.~(\ref{amp-jpsiV}).

\subsection{Decay into $\bm{J/\psi \, \pi^+ \pi^-}$}

We assume that the decay of $X$ into $J/\psi \, \pi^+ \pi^-$
proceeds through a transition of $X$ to $J/\psi \, \rho$.
The T-matrix element is then given in terms of the 
unknown coupling constant $G_{X \psi \rho}$
by Eqs.~(\ref{T-psipi}) and (\ref{amp-jpsiV}) with $V = \rho$.
The expression for the amplitude ${\cal A}_\nu$ for 
$\rho \to \pi^+ \pi^-$ is given in Eq.~(\ref{amp-rho}).
We obtain the decay rate by squaring the amplitude, 
summing over spins, and integrating over phase space.
The differential decay rate into $J/\psi \, \pi^+ \pi^-$
as a function of the invariant mass $Q$ of the two pions is
%-----------------
\begin{eqnarray}
{d \Gamma \over d Q}[X \to J/\psi \, \pi^+ \pi^-]
&=&  {|G_{X \psi \rho}|^2 G_{v \pi \pi}^2 
	\over 9216  \pi^3 m_X^5 m_\psi^2} 
{(Q^2 - 4 m_\pi^2)^{3/2} \lambda^{1/2}(m_X,m_\psi,Q) 
	\over (Q^2 - m_\rho^2)^2 + m_\rho^2 \Gamma_\rho^2}
\nonumber
\\ 
&& \hspace{-3cm}
\times
\left[ (m_X^2 + m_\psi^2) (m_X^2 - m_\psi^2)^2
- 2 (m_X^4 - 4 m_X^2 m_\psi^2 + m_\psi^4) Q^2 
+ (m_X^2 + m_\psi^2) Q^4 \right] ,
\label{dGam-dQ:rho}
\end{eqnarray}
%-----------------
where $\lambda(x,y,z)$ is the triangle function:
\begin{eqnarray}
\lambda(x,y,z)=x^4+y^4+z^4-2(x^2y^2+y^2z^2+z^2x^2).
\end{eqnarray}
After integrating over the pion invariant mass, 
the decay rate is
%-----------------
\begin{eqnarray}
\Gamma[X\to J/\psi \, \pi^+\pi^-]
&=& |G_{X \psi \rho}|^2 \,  ( 223  \, {\rm keV} ).
\label{Gam-psi2pi}   
\end{eqnarray}
%----------------- 
 
\begin{figure}[t]
\includegraphics[width=7.8cm]{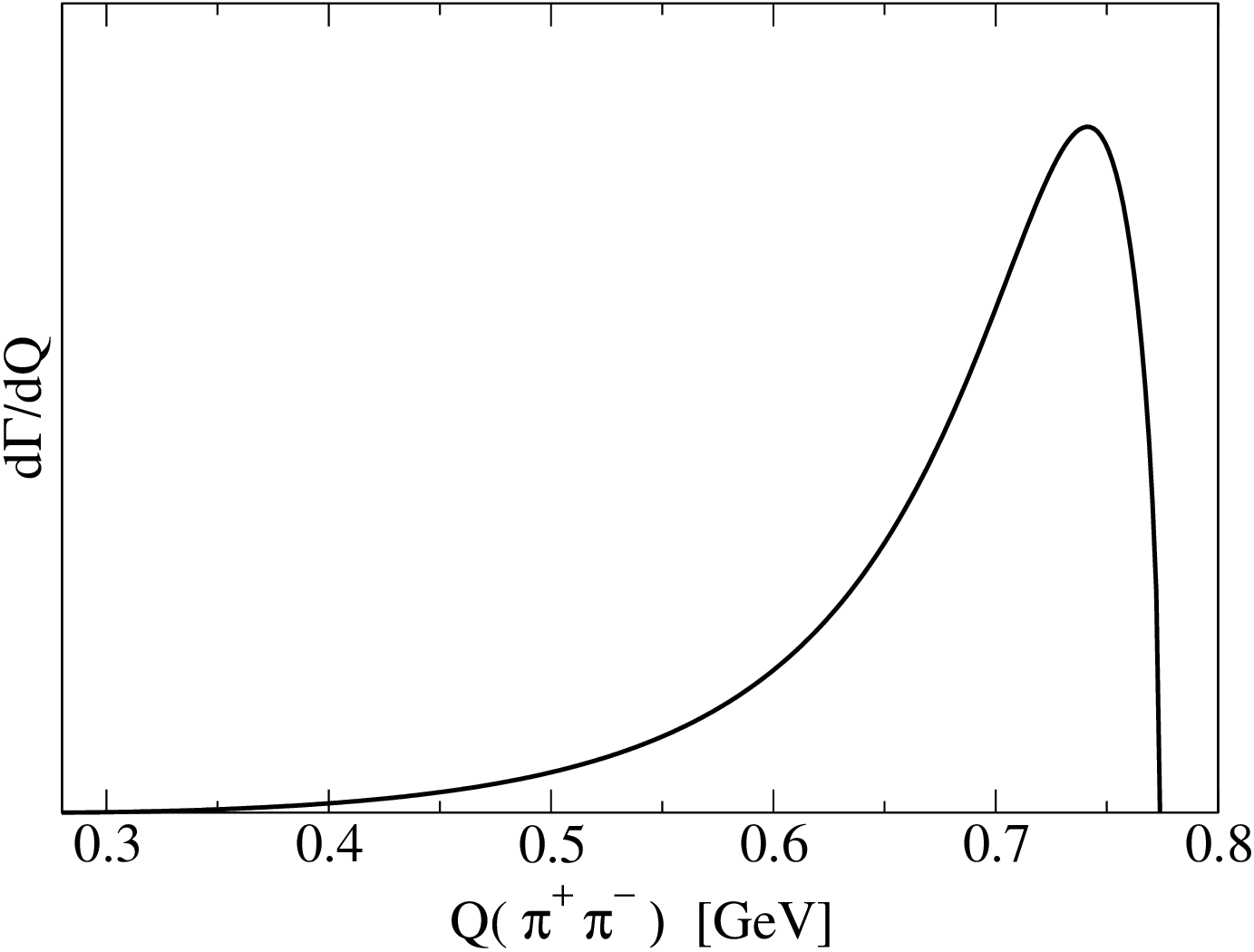}
\caption{Invariant mass distributions 
for the pions in the decay $X \to J/\psi \, \pi^+\pi^-$.
\label{fig:plotrho}}
\end{figure}

The shape of the pion invariant mass distribution 
for the decay of $X$ into $J/\psi \, \pi^+\pi^-$
is shown in Fig.~\ref{fig:plotrho}.
Its qualitative features 
are dominated by the phase space factor $\lambda^{1/2}(m_X,m_\psi,Q)$, 
which cuts the distribution off at the endpoint $Q = m_X - m_{\psi}$,
and the vector meson resonance factor, which has its maximum at 
$Q = m_\rho$  just outside the kinematic region.  
Most of the support for $d \Gamma/d Q$
comes from within $\Gamma_\rho$ of the upper endpoint.

%------------------------------------------------------------------------------------
%------------------------------------------------------------------------------------

\subsection{Decay into $\bm{J/\psi \, \pi^+ \pi^- \pi^0}$}

We assume that the decay of $X$ into $J/\psi \, \pi^+ \pi^- \pi^0$
proceeds through a transition of $X$ to $J/\psi \, \omega$.
The T-matrix element is then given in terms of the 
unknown coupling constant $G_{X \psi \omega}$
by Eqs.~(\ref{T-psipi}) and (\ref{amp-jpsiV}) with $V = \omega$.
The expression for the amplitude for $\omega \to 3 \pi$ is given 
in Eq.~(\ref{amp-omega}).
We obtain the decay rate by squaring the amplitude, 
summing over spins, and integrating over phase space.
The differential decay rate into $J/\psi \pi^+ \pi^- \pi^0$ 
as a function of the invariant mass $Q$ 
of the 3 pions can be reduced to a 2-dimensional integral:
%-----------------
\begin{eqnarray}
{d \Gamma \over d Q}[X \to J/\psi \, \pi^+ \pi^- \pi^0]
&=& { |G_{X \psi \omega}|^2 (\cos\theta_v+\sqrt{2}\sin\theta_v)^2
     \over 3072 \pi^5 m_X^5 m_\psi^2 F_{\pi}^6}
  { \lambda^{1/2}(m_X,m_\psi,Q)
        \over Q [(Q^2 - m_\omega^2)^2 + m_\omega^2 \Gamma_\omega^2] }
\nonumber
\\
&& \hspace{-3cm}\times
\left[ (m_X^2 + m_\psi^2) (m_X^2 - m_\psi^2)^2
- 2 (m_X^4 - 4 m_X^2 m_\psi^2 + m_\psi^4) Q^2 
+ (m_X^2 + m_\psi^2) Q^4 \right]
\nonumber
\\
&& \hspace{-3cm}\times
\int d s_{12}
\int d s_{23}
\left[s_{12} \, s_{23} \, s_{31} - m_{\pi}^2 (Q^2-m_{\pi}^2)^2
   \right]
\nonumber
\\
&& \hspace{-2cm}\times
\left| C_{v 3\pi}
+ \frac{ G_{v \pi \pi}C_{vv\pi}
F_{\pi}^2 } { m_v^2 }
\left( 1 -  \mbox{$1\over 3$} \,
\left[f_\rho(s_{12})+f_\rho(s_{23})+f_\rho(s_{31})\right]
 \right) \right|^2 ,
\label{dGam-dQ:omega}
\end{eqnarray}
%-----------------
where $s_{12}$, $s_{23}$, and $s_{31}$ are the squares of the invariant 
masses of the three pairs of pions.
We have suppressed the limits of integration in the integrals over
$s_{12}$ and $s_{23}$.  After integrating over the pion invariant masses, 
the decay rate is
%-----------------
\begin{eqnarray}
\Gamma[X\to J/\psi \, \pi^+\pi^-\pi^0]
&=& |G_{X \psi \omega}|^2 ( 19.4  \, {\rm keV} ).
\label{Gam-psi3pi} 
\end{eqnarray}
%-----------------

\begin{figure}[t]
\includegraphics[width=7.8cm]{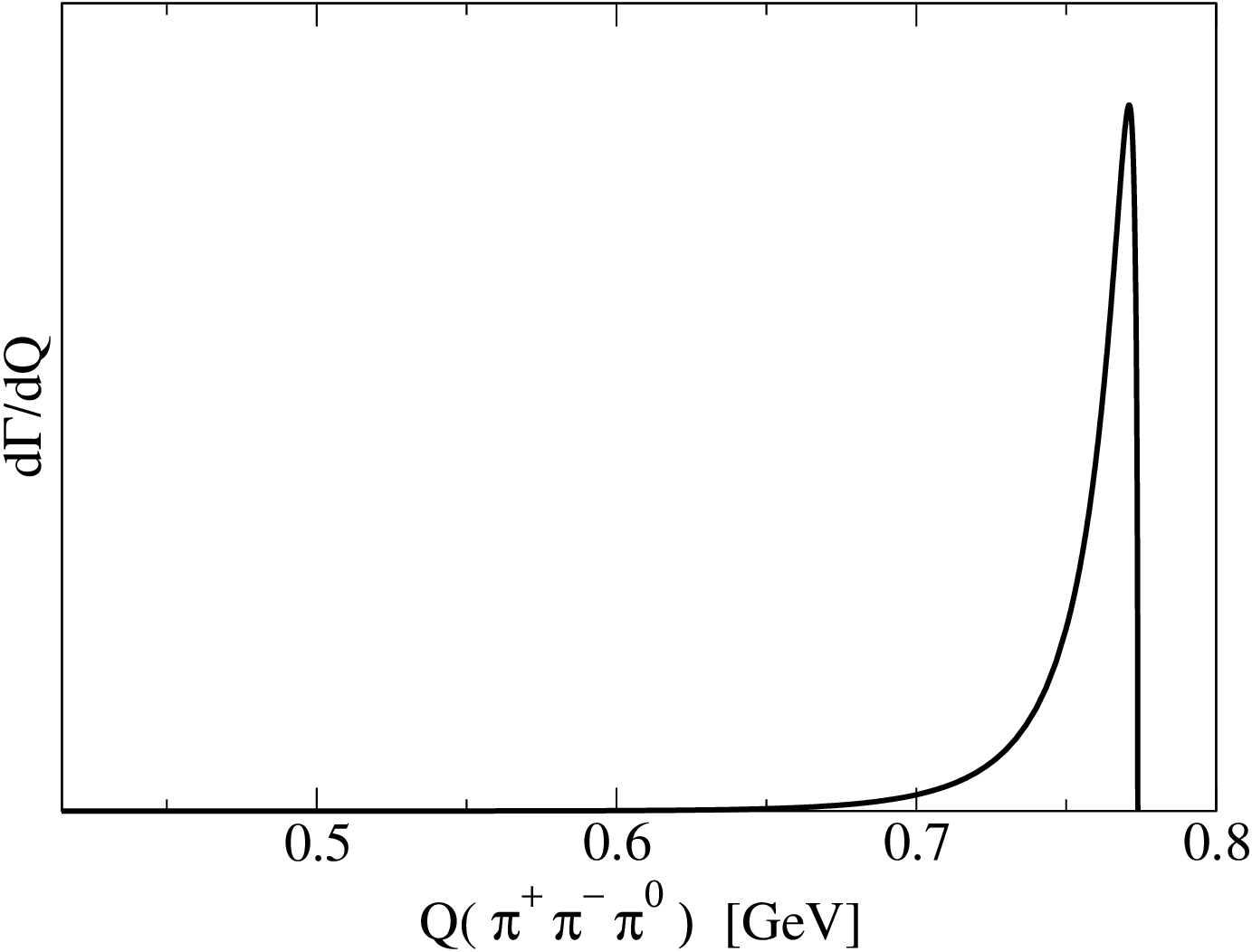}
\caption{Invariant mass distributions for the pions 
in the decays $X \to J/\psi \, \pi^+\pi^-\pi^0$.
\label{fig:plot}}
\end{figure}

The shape of the pion invariant mass distributions 
for the decay of $X$ into $J/\psi \, \pi^+\pi^-\pi^0$ is shown 
in Fig.~\ref{fig:plot}.
Its qualitative features 
are dominated by the phase space factor $\lambda^{1/2}(m_X,m_\psi,Q)$, 
which cuts the distribution off at the endpoint $Q = m_X - m_{\psi}$,
and the vector meson resonance factor, which has its maximum at 
$Q = m_\omega$  just outside the kinematic region.  
Most of the support for $d \Gamma/d Q$
comes from within a few widths $\Gamma_\omega$ of the upper endpoint.

%------------------------------------------------------------------------------------
%------------------------------------------------------------------------------------

The ratio of the decay rates in Eqs.~(\ref{Gam-psi2pi}) 
and (\ref{Gam-psi3pi}) is 
%-----------------
\begin{eqnarray}
 \frac{\Gamma[X\to J/\psi \, \pi^+\pi^-\pi^0]} 
   {\Gamma[X\to J/\psi \, \pi^+\pi^-]}
   = 0.0870 \; \frac{|G_{X \psi \omega}|^2}{|G_{X \psi \rho}|^2}.
\label{ratio_psipi}
\end{eqnarray}
%-----------------
By comparing this to Belle's result in Eq.~(\ref{BR:pipipi/pipi})
for the ratio of the branching fractions, 
we can obtain an estimate of the ratio of the coupling constants:
%-----------------
\begin{eqnarray}
{|G_{X \psi \omega}|^2 \over |G_{X \psi \rho}|^2} \approx 11.5 \pm 5.7.
\label{Gomega/Grho}
\end{eqnarray}
%-----------------

\subsection{Decay into $\bm{J/\psi \, \pi^0 \gamma}$}

We assume that the decay of $X$ into $J/\psi \, \pi^0 \gamma$
proceeds through transitions of $X$ to $J/\psi \, \rho$ and 
$J/\psi \, \omega$.
The T-matrix element is then given by Eq.~(\ref{T-psipi})
summed over $V = \rho, \omega$.
The amplitudes for $V\to \pi^0 \gamma$ are given in 
Eqs.~(\ref{amp:pigamma}).
The differential decay rate with respect to the invariant mass $Q$ 
of the $\pi^0 \gamma$ is  
%-----------------
\begin{eqnarray}
{d\Gamma \over dQ}[X \to J/\psi \,\pi^0 \gamma]
&=& {\alpha_{\rm em}  
 \left( C_{v \pi \gamma}
+ G_{v \gamma} C_{vv\pi} F_\pi^2 / m_v^2 \right)^2
\over 648 \pi^2 m_X^5 m_\psi^2 F_\pi^2} \,
{(Q^2 - m_\pi^2)^3 \lambda^{1/2}(m_X,m_\psi,Q) \over Q}
\nonumber
\\
&& \hspace{-3cm} \times 
\left[ (m_X^2 + m_\psi^2) (m_X^2 - m_\psi^2)^2
- 2 (m_X^4 - 4 m_X^2 m_\psi^2 + m_\psi^4) Q^2 
+ (m_X^2 + m_\psi^2) Q^4 \right]
\nonumber
\\
&& \hspace{-3cm} \times 
\left| {G_{X \psi \rho} 
	\over Q^2 - m_\rho^2 + i m_\rho \Gamma_\rho}
+ {G_{X \psi \omega} \sqrt{3} (\cos \theta_v + \sqrt{2}\sin\theta_v)
	\over Q^2 - m_\omega^2 + i m_\omega \Gamma_\omega} \right|^2 .
\end{eqnarray}
%-----------------
After integrating over the $\pi^0 \gamma$ invariant mass,
the decay rate is
%-----------------
\begin{eqnarray}
\Gamma[X \to J/\psi \, \pi^0 \gamma]
&=& \big[ |G_{X \psi \omega}|^2 + 0.026 \, |G_{X \psi \rho}|^2 
\nonumber
\\
&& 
+ (0.163 \cos\phi + 0.215 \sin \phi)
|G_{X \psi \omega}| |G_{X \psi \rho}| \big]
( 3.24 \, {\rm keV} ),
\label{Gam-psipigamma}
\end{eqnarray}
%-----------------
where $\exp(i \phi)$ is the relative phase between $G_{X \psi \omega}$
and $G_{X \psi \rho}$.  The estimate of the ratio 
$|G_{X \psi \omega}|^2/|G_{X \psi \rho}|^2$
in Eq.~(\ref{Gomega/Grho}) suggests that the $|G_{X \psi \omega}|^2$
term in Eq.~(\ref{Gam-psipigamma}) dominates. 
If this is the case, the branching fraction 
for the decay of $X$ into $J/\psi \, \pi^0 \gamma$ should be smaller 
than that for $J/\psi \, \pi^+ \pi^- \pi^0$
by a factor of about 0.17.

\subsection{Decay into $\bm{J/\psi \, \gamma}$}

Having chosen the transition amplitude in Eq.~(\ref{amp-jpsiV})
so that it satisfies $Q^\mu {\cal A}_\mu=0$,
we can use vector meson dominance to calculate the  
partial width for the decay of $X$ into $J/\psi \, \gamma$.  
The T-matrix element is 
%-----------------
\begin{eqnarray}
{\cal T}[X \to J/\psi \, \gamma] &=&
G_{v\gamma} F_\pi^2 e 
\left( {G_{X \psi \rho} 
	\over m_\rho^2 - i m_\rho \Gamma_\rho}
+ {G_{X \psi \omega} \cos \theta_v / \sqrt{3}
	\over m_\omega^2 - i m_\omega \Gamma_\omega} \right)
\, \varepsilon_{\mu\nu\alpha\beta} Q^{\mu} \epsilon_X^{\nu} 
	{\epsilon_{\psi}^{\alpha}}^* {\epsilon_{\gamma}^{\beta}}^*,
\label{T-psigamma}
\end{eqnarray}
%-----------------
where $Q$ is the 4-momentum of the photon.
The value of the coupling constant $G_{v\gamma}$ is given in 
Eq.~(\ref{Gvgam}).  The result for the decay rate is
%-----------------
\begin{eqnarray}
\Gamma[X \to J/\psi \, \gamma] &=&
{\alpha_{\rm em} G_{v\gamma}^2 F_\pi^4 
	(m_X^2 + m_\psi^2) (m_X^2 - m_\psi^2)^3
	\over 24 m_X^5 m_\psi^2}
\nonumber
\\
&& \times \left| {G_{X \psi \rho} 
	\over m_\rho^2 - i m_\rho \Gamma_\rho}
+ {G_{X \psi \omega} \cos \theta_v / \sqrt{3}
	\over m_\omega^2 - i m_\omega \Gamma_\omega} \right|^2 .
\label{rate-psigamma}
\end{eqnarray}
%-----------------
If the widths in the vector meson propagators are neglected
and if we use $m_\rho \approx m_\omega$,
the decay rate in Eq.~(\ref{rate-psigamma}) reduces to
%-----------------
\begin{eqnarray}
\Gamma[X \to J/\psi \, \gamma] =
 |G_{X \psi \rho} + 0.30 \, G_{X \psi \omega} |^2
( 5.51  \, {\rm keV} ).
\label{Gam-psigamma}
\end{eqnarray}
%-----------------
Our estimate in Eq.~(\ref{Gomega/Grho}) implies that 
$|G_{X \psi \omega}|$ is much larger than $|G_{X \psi \rho}|$.
However, the larger magnitude of $G_{X \psi \omega}$
is compensated by the vector meson mixing factor 
$\cos \theta_v/\sqrt{3} = 0.30$,
so the $G_{X \psi \rho}$ and $G_{X \psi \omega}$ terms may be equally 
important. 
Using the partial widths in Eqs.~(\ref{Gam-psi2pi}) and (\ref{Gam-psi3pi}),
we can relate the branching fractions for $J/\psi \, \gamma$ to those for 
$J/\psi \, \pi^+ \pi^-$ and $J/\psi \, \pi^+ \pi^- \pi^0$:
%-----------------
\begin{eqnarray}
{\rm Br}[X \to J/\psi \, \gamma] &=& 
0.025 \, {\rm Br}[X \to J/\psi \, \pi^+ \pi^-] 
+ 0.026 \, {\rm Br}[X \to J/\psi \, \pi^+ \pi^- \pi^0]
\nonumber
\\
&& + 0.050 \cos \phi \, \left( {\rm Br}[X \to J/\psi \, \pi^+ \pi^-] \,
		{\rm Br}[X \to J/\psi \, \pi^+ \pi^- \pi^0] \right)^{1/2} ,
\end{eqnarray}
%-----------------
where $\exp(i \phi)$ is the relative phase between $G_{X \psi \omega}$
and $G_{X \psi \rho}$.  This prediction is compatible with the 
measurements of the branching ratios in Eqs.~(\ref{BR:pipipi/pipi}) 
and (\ref{BR:gamma/pipi}) if the angle $\phi$ is small.

\subsection{Factorization of short-distance decay rates}

The decay modes of the $X(3872)$ can be classified into 
{\it long-distance decays} and {\it short-distance decays}.
The long-distance decay modes are $D^0 \bar D^0 \pi^0$ and  
$D^0 \bar D^0 \gamma$, which proceed through the decay of a 
constituent $D^{*0}$ or $\bar D^{*0}$.  These decays are dominated 
by a component of the wavefunction of the $X$ in which the separation 
of the $D$ and $D^*$ is of order $|a|$.  
These long-distance decays involve interesting interference effects 
between the $D^0 \bar D^{*0}$ and $D^{*0} \bar D^0$ components 
of the wavefunction \cite{Voloshin:2003nt}.
The short-distance decays are dominated by a component of the 
wavefunction in which the separation of the $D$ and $D^*$ 
is of order $m_\pi$ or smaller.  Examples are the observed decay modes
$J/\psi \, \pi^+ \pi^-$, $J/\psi \, \pi^+ \pi^- \pi^0$, 
and $J/\psi \, \gamma$.

Short-distance decays of the $X$ into a hadronic final state $H$
involve well-separated momentum scales.
The $D D^*$ wavefunction of the $X$ involves the momentum scale $1/|a|$  
set by the large scattering length.  The transition of the $D D^*$ 
to $H$ involves momentum scales $m_{\pi}$ and larger.
The separation of scales $|a| \gg 1/m_\pi$ can be exploited
by using a factorization formula for the decay rate \cite{Braaten:2005jj}.  
In limit $|a| \gg 1/m_\pi$, the leading term in the T-matrix element 
for the decay $X \to H$ can be separated into 
a short-distance factor and a long-distance factor:
%-----------------
\begin{eqnarray}
{\cal T}[ X \to H] = 
{\cal A}_{\rm short} [X \to H] \times {\cal A}_X.
\label{XtoH:fact}
\end{eqnarray}
%-----------------
The short-distance factor ${\cal A}_{\rm short}$ in Eq.~(\ref{XtoH:fact}) 
has a well-behaved limit as $|a| \to \infty$.  
The long-distance factor ${\cal A}_X$ is the universal amplitude 
given in Eq.~(\ref{AX-uni}).  If the complex scattering length is 
parameterized as in Eq.~(\ref{a-complex}), this factor is
%-----------------
\begin{eqnarray}
{\cal A}_X = \left( \sqrt{2 \pi}/\mu \right) 
\left( \gamma_{\rm re} + i \gamma_{\rm im} \right)^{1/2}.
\label{AX-complex}
\end{eqnarray}
%-----------------
When applied to decays of 
$X$ into $J/\psi$ and light hadrons, the factorization formula in
Eq.~(\ref{XtoH:fact}) implies that the coupling constants
$G_{X \psi \rho}$ and $G_{X \psi \omega}$ have a long-distance factor
${\cal A}_X$.

The factorization formula for the T-matrix element in Eq.~(\ref{XtoH:fact}) 
implies a factorization formula for the decay rate:
%-----------------
\begin{eqnarray}
\Gamma[X \to H] = 
\Gamma_{\rm short}[X \to H] \times |{\cal A}_X|^2 .
\label{Gamma:fact}
\end{eqnarray}
%-----------------
The short-distance factor $\Gamma_{\rm short}$ in Eq.~(\ref{Gamma:fact}) 
has a well-behaved limit as $|a| \to \infty$.  
Using the expressions in Eqs.~(\ref{E-bs}) and (\ref{Gam-bs}) for the 
binding energy and the width of the molecule, the long-distance factor 
in Eq.~(\ref{Gamma:fact}) can be expressed as
%-----------------
\begin{eqnarray}
|{\cal A}_X|^2 = 
\sqrt{8 \pi^2 / \mu^3} \left[ E_X + \Gamma_X^2/(16 E_X) \right]^{1/2}.
\label{AXsq}
\end{eqnarray}
%-----------------

Predictions for the rates for short-distance decays of the $X$
can be obtained from models for low-energy hadrons in which 
the parameters have been tuned to obtain a small binding energy 
$E_X$, such as Swanson's model \cite{Swanson:2003tb}. 
In such models, calculations using the most straightforward 
numerical methods tend to become increasingly unstable as the 
binding energy is tuned toward 0, because the small binding energy 
results from a delicate cancellation. 
The factorization formula in Eqs.~(\ref{Gamma:fact}) 
and (\ref{AXsq}) can be useful 
for extrapolating the predictions of a model to other values of 
the binding energy $E_X$.  In many models, it is difficult to take 
into account effects of the width $\Gamma_X$ of the molecule.
Given the prediction of a model in which the width has been neglected,
the factorization formula in Eqs.~(\ref{Gamma:fact}) and (\ref{AXsq}) 
can be used to
take into account the nonzero width $\Gamma_X$ consistently.

In order to use the factorization formula in Eqs.~(\ref{Gamma:fact})
and (\ref{AXsq}) to extrapolate a partial width calculated using a model 
to other values of the binding energy 
and the width, the calculation must be carried out 
for small enough binding energy that the model is in the 
universal scaling regime where observables scale as powers 
of the binding energy.
For example, the probabilities for components of the wavefunction 
other than $D^0 \bar D^{*0}$ and $D^{*0} \bar D^{0}$
should scale as $E_X^{1/2}$.
In Swanson's model with $E_X = 0.7$ MeV, this universal scaling
behavior is satisfied by the probability $Z_{\psi \omega}$
but not by $Z_{\psi \rho}$, as is evident in Fig.~\ref{fig:matching}.
The delayed onset of the universal behavior for the probability 
$Z_{\psi \rho}$ can perhaps be attributed to the weaker coupling 
of $X$ to isospin-1 states.
In the next section, we will use Swanson's result for 
$Z_{\psi \omega}$ to estimate the coupling constant $G_{X \psi \omega}$.

%\newpage

\section{Partial width for $\bm{X \to J/\psi \, \pi^+ \pi^- \pi^0}$} 
\label{sec:model}

The partial widths of the $X$ calculated in Section~\ref{sec:Xdecay}
are expressed in terms of unknown coupling constants 
$G_{X \psi \rho}$ and $G_{X \psi \omega}$.  In this section, 
we use a simple 2-channel scattering model to show that 
$|G_{X \psi \omega}|$ can be deduced from the probability 
$Z_{\psi \omega}$ for the $J/\psi \, \omega$ component of the $X$.
We then use the probability $Z_{\psi \omega}$ in Swanson's
model to give a quantitative prediction for the partial width 
for the $X$ to decay into $J/\psi \, \pi^+ \pi^- \pi^0$. 

\subsection{Two-channel scattering model}
\label{sec:2chmodel}

Cohen, Gelman, and van Kolck have constructed a renormalizable 
effective field theory that describes two scattering channels 
with S-wave contact interactions \cite{Cohen:2004kf}.
We will refer to this model as the {\it two-channel scattering model}.
An essentially equivalent model has been used to describe the effects 
of $\Delta \Delta$ states on the two-nucleon system \cite{Savage:1996tb}.
The parameters of this model can be tuned to produce a large 
scattering length in the lower energy channel.  It can be used as
a simple model for the effects on the $D^0 \bar D^{*0}$/$D^{*0} \bar D^0$
system of other hadronic channels with nearby thresholds, 
such as $J/\psi \, \rho$ and $J/\psi \, \omega$.

The two-channel model of Ref.~\cite{Cohen:2004kf} 
describes two scattering 
channels with S-wave contact interactions only.
We label the particles in the first channel $1a$ and $1b$
and those in the second channel $2a$ and $2b$.
We denote the reduced masses in the two channels by $\mu$ and $\mu_2$.
Renormalized observables in the 2-body sector are expressed in terms of 
4 parameters: three interaction parameters $a_{11}$, $a_{22}$, 
and $a_{12}=a_{21}$ with dimensions of length and the
energy gap $\Delta$ between the two scattering channels,
which is determined by the masses of the particles:
%-----------------
\begin{eqnarray}
\Delta &=& m_{2a} + m_{2b}- (m_{1a}+m_{1b}).
\label{Delta-real}
\end{eqnarray} 
%-----------------
The scattering parameters in Ref.~\cite{Cohen:2004kf} were defined
in such a way that $a_{11}$ and $a_{22}$ reduce in the limit 
$a_{12} \to \pm \infty$ to the scattering lengths for the two channels.  
The truncated connected transition amplitude 
${\cal A}(E)$ for this coupled-channel system is 
a $2\times 2$ matrix that depends on the energy $E$ 
in the center-of-mass frame. 
If that energy is measured relative to the threshold $m_{1a}+m_{1b}$ 
for the first scattering channel, 
the inverse of the matrix ${\cal A}(E)$ is
%-----------------
\begin{eqnarray}
{\cal A}(E)^{-1}= \frac{1}{2\pi}
\begin{pmatrix}
\mu \big[ -1/a_{11} +\sqrt{- 2 \mu E} \, \big]  & 
	\sqrt{\mu \mu_2}/a_{12}   \\
\sqrt{\mu \mu_2}/a_{12} & 
	\mu_2 \big[ -1/a_{22}  + \sqrt{2 \mu_2 (\Delta - E)} \, \big]
\end{pmatrix}. 
\label{A-2cm}
\end{eqnarray}
%-----------------
The square roots are defined for negative real arguments by the 
prescription $E \to E + i \epsilon$ with $\epsilon \to 0^+$.
The explicit expressions for the $11$ and $12$ entries
of this matrix are
\begin{subequations}
\begin{eqnarray}
{\cal A}_{11}(E) &=& {2 \pi \over \mu} 
\left( - {1 \over a_{11}} + \sqrt{-2\mu E} 
- {1 \over a_{12}^2} 
\Big[ -1/a_{22} + \sqrt{2 \mu_2 (\Delta -  E )} \, \Big]^{-1}  
\right)^{-1},
\label{A11-2cm}
\\
{\cal A}_{12}(E) &=& {2 \pi \over \sqrt{\mu \mu_2}} 
\left( {1 \over a_{12}}
- a_{12} \left[ - {1 \over a_{11}} + \sqrt{-2\mu E } \right]  
	\left[ - {1 \over a_{22}} + \sqrt{2\mu_2 (\Delta - E )} \right] 
\right)^{-1}. 
\label{A12-2cm}
\end{eqnarray}
\end{subequations}
The amplitudes defined by (\ref{A-2cm}) are for transitions between states 
with the standard nonrelativistic normalizations.  The transitions 
between states with the standard relativistic normalizations
are obtained by multiplying by a factor $\sqrt{2 m_i}$ for every 
particle in the initial and final state.
The T-matrix element $T_{11}(p)$ for the elastic scattering of particles 
in the first channel with relative momentum $p$ is obtained by evaluating
${\cal A}_{11}(E)$ at the energy $E = p^2/(2 \mu)$.
The scattering length is determined by the T-matrix element at $p=0$:
$T_{11}(0) = - 2 \pi a/\mu$.
The inverse scattering length $1/a$ is therefore
%-----------------
\begin{eqnarray}
{1 \over a} &=& \frac{1}{a_{11}}
+ \frac{1}{a_{12}^2} \Big[ \sqrt{2\mu_2 \Delta} - 1/a_{22} \, \Big]^{-1} .
\label{a-2cm}
\end{eqnarray}
%-----------------
%

If the matrix ${\cal A}(E)$ given by Eq.~(\ref{A-2cm}) 
has a pole on the physical sheet at $E = -\kappa^2/(2\mu)$,
there is a bound state below the scattering threshold 
for the first channel with binding energy $E_X =\kappa^2/(2\mu)$.
The binding momentum $\kappa$ satisfies
%-----------------
\begin{eqnarray}
\kappa = \frac{1}{a_{11}}
+ \frac{1}{a_{12}^2} 
\Big[ - 1/a_{22} +\sqrt{2\mu_2 \Delta + (\mu_2/\mu) \kappa^2} \, \Big]^{-1}.
\label{kappa-2cm}
\end{eqnarray}
%-----------------
The momentum-space wavefunction $\psi(p)$
for the bound state is a column vector whose two components
are the amplitudes for the bound state to consist of particles 
with relative momentum $p$ in the first and second channel, respectively.
The wavefunction can be deduced from 
the behavior of ${\cal A}(E)$ near the bound-state pole:
%-----------------
\begin{eqnarray}
{\cal A}(E) \longrightarrow - {1 \over E + \kappa^2/(2\mu)} 
	\begin{pmatrix} {\cal A}_{X1}\\ {\cal A}_{X2} \end{pmatrix}
  \otimes \begin{pmatrix} {\cal A}_{X1} & {\cal A}_{X2} \end{pmatrix}.
\end{eqnarray}
%-----------------
The components ${\cal A}_{X1}$ and ${\cal A}_{X2}$ of the column vector are the amplitudes 
for transitions from the bound state to particles in the first and 
second channels, respectively.
They satisfy
%-----------------
\begin{eqnarray}
\mu [-1/a_{11} + \kappa ] \, {\cal A}_{X1}
+ [\sqrt{\mu \mu_2}/a_{12}] \, {\cal A}_{X2} = 0.
\end{eqnarray}
%-----------------
Because the only interactions in the two-channel model are contact 
interactions, the dependence of the wavefunction on the relative 
momentum of the constituents comes only from propagators.
The wavefunction can be expressed in the form
%-----------------
\begin{eqnarray}
\psi(p) = N
\begin{pmatrix}
	2 \mu {\cal A}_{X1}[p^2 + \kappa^2]^{-1} \\
	2 \mu_2 {\cal A}_{X2} [p^2 + 2 \mu_2 \Delta + (\mu_2/\mu)\kappa^2]^{-1}
\end{pmatrix} ,
\label{psi-2cm}
\end{eqnarray}
%-----------------
where $N$ is a normalization constant.
The normalization condition
%-----------------
\begin{eqnarray}
\int {d^3 p \over (2 \pi)^3} \left( |\psi_1(p)|^2 + |\psi_2(p)|^2 \right)
= 1
\label{psi-norm}
\end{eqnarray}
%-----------------
can be expressed as $Z_1 + Z_2 = 1$, where $Z_1$ and $Z_2$ 
are the probabilities for the bound state to consist of the particles 
in the first and second channels, respectively.  
The probability $Z_1$ for the first channel is given by
%-----------------
\begin{eqnarray}
Z_1^{-1} = 1 + {(\mu_2/\mu) a_{12}^2 (-1/a_{11} + \kappa)^2 \kappa 
	\over \sqrt{2 \mu_2 \Delta + (\mu_2/\mu)\kappa^2}}.
\label{Z-2cm}
\end{eqnarray}
%-----------------

\subsection{Two-channel model with large scattering length}

In the two-channel model of Ref.~\cite{Cohen:2004kf},
a large scattering length $a$ in the first channel can be obtained
by fine-tuning the parameters $a_{11}$, $a_{22}$, $a_{12}$, and $\Delta$.
The natural momentum scale $\Lambda$ associated with low-energy 
elastic scattering in the first channel is set by the magnitudes of
$a_{11}^{-1}$, $a_{22}^{-1}$, $a_{12}^{-1}$, and $(2 \mu_2 \Delta)^{1/2}$.
There are various ways to tune the parameters so that $|a|$ 
is large compared to $\Lambda^{-1}$.  
For example, $a$ can be tuned to $\pm \infty$ by tuning the 
scattering parameter $a_{11}$ to the critical value
$- a_{12}^2 [ \sqrt{2\mu_2 \Delta} - 1/a_{22} ]$.

As $a$ is tuned to be much larger than the natural momentum scale, 
the amplitude ${\cal A}_{11}(E)$
for $|E| \ll \Lambda^2/(2\mu)$ approaches the
universal expression given in Eq.~(\ref{A-uni}). 
The solution to Eq.~(\ref{kappa-2cm})
for the binding momentum $\kappa$ approaches $1/a$, so if $a>0$, 
there is a bound state with the universal binding energy
in Eq.~(\ref{Eb}).  The first component of the wavefunction 
in Eq.~(\ref{psi-2cm}) approaches the universal expression in
Eq.~(\ref{psi-uni}), while the probability of the second component 
approaches 0 as $1/a$.
The amplitude ${\cal A}_{X1}$ for the transition from the bound state
to particles in the first channel also approaches the universal 
amplitude ${\cal A}_X$ in Eq.~(\ref{AX-uni}).

There are also universal features associated with transitions 
from the bound state to particles in the second channel.
If $|a| \gg \Lambda^{-1}$, the leading term in the amplitude 
for the transition of the 
weakly-bound state $X$ to particles in the second channel is 
\begin{eqnarray}
{\cal A}_{X2} = 
- {\sqrt{\mu/\mu_2}  \over a_{12}}
\Big[ \sqrt{2\mu_2 \Delta} - 1/a_{22} \, \Big]^{-1} 
{\cal A}_X ,
\label{AX2-largea}
\end{eqnarray}
where ${\cal A}_X$ is the universal amplitude given in Eq.~(\ref{AX-uni}).
This equation is a factorization formula that expresses the transition 
amplitude as the product of a short-distance factor and the universal
long-distance factor ${\cal A}_X$.
Using Eq.~(\ref{Z-2cm}), the probability $Z_2 = 1 -Z_1$ for the bound state 
to consist of particles in the second channel reduces to 
%-----------------
\begin{eqnarray}
Z_2 = 
{(\mu_2/\mu)  \over a_{12}^2 \sqrt{2\mu_2 \Delta}}
\Big[ \sqrt{2\mu_2 \Delta} - 1/a_{22} \, \Big]^{-2} 
{1 \over a} .
\end{eqnarray}
%-----------------
Note that the probability $Z_2$ differs from 
$|{\cal A}_{X2}|^2$ only by kinematic factors:
%-----------------
\begin{eqnarray}
|{\cal A}_{X2}|^2 = \sqrt{8 \pi^2 \Delta/\mu_2^3} \, Z_2 .
\label{A-Z}
\end{eqnarray}
%-----------------
This relation also follows directly from the wavefunction
in Eq.~(\ref{psi-2cm}) if we use the fact that the normalization 
factor $N$ approaches 1 as $a\to \infty$.  Thus the relation 
between the probability and the transition amplitude 
in Eq.~(\ref{A-Z}) is not specific to the 2-channel model.  
It applies more generally to any 2-particle component 
of the bound state whose wavefunction can be approximated by
$(p^2 + 2 \mu_2 \Delta)^{-1}$, where $\Delta$ is the energy gap.
It requires only that $\Delta$ is small enough that 
the interaction in that channel can be approximated by an 
S-wave contact interaction at momenta comparable to 
$\sqrt{2 \mu_2 \Delta}$.

\subsection{Partial width into $\bm{J/\psi \,\pi^+\pi^-\pi^0}$}

We can use results from Swanson's model to estimate 
$|G_{X\psi \omega}|$, thereby determining the unknown constant
in the expression in Eq.~(\ref{Gam-psi3pi}) for the partial width 
for $X \to  J/\psi \, \pi^+ \pi^- \pi^0$.   
The relativistic amplitude for the transition 
from $X$ to $J/\psi \, \omega$ is given by the contraction of the 
amplitude ${\cal A}_{\mu}$ in Eq.~(\ref{amp-jpsiV}) with a 
polarization vector for the $\omega$.  The corresponding 
nonrelativistic amplitude ${\cal A}_{X\psi \omega}$  
is the analog of the transition amplitude ${\cal A}_{X2}$ in 
Eq.~(\ref{AX2-largea}) for the 2-channel model.
In the rest frame of the $X$, the relativistic amplitude
differs from the nonrelativistic amplitude 
by a factor of $\sqrt{2m_i}$ for every external particle:
%-----------------
\begin{eqnarray}
{\epsilon_\omega^\mu}^* {\cal A}_\mu[X \to J/\psi \, \omega] = 
(8 m_X m_\psi m_\omega)^{1/2} {\cal A}_{X\psi \omega} .
\label{A-A}
\end{eqnarray}
%-----------------
Using the expression for the amplitude ${\cal A}_\mu$ in 
Eq.~(\ref{amp-jpsiV}) and the fact that the rest frame of the $X$ 
is almost identical to that of the $\omega$, 
the left side of Eq.~(\ref{A-A}) is 
%-----------------
\begin{eqnarray}
{\epsilon_\omega^\mu}^* {\cal A}_\mu[X \to J/\psi \, \omega] =
G_{X\psi \omega} m_\omega 
\bm{\epsilon}_X \cdot (\bm{\epsilon}_\psi \times \bm{\epsilon}_\omega)^* .
\label{G-A}
\end{eqnarray}
%-----------------
The transition amplitude ${\cal A}_{X\psi \omega}$ on the right side 
of Eq.~(\ref{A-A}) must have the same dependence on the polarization 
vectors of the $J/\psi$ and $\omega$.  There are two independent 
pairs of spin states for $J/\psi$ and $\omega$ that couple to 
any given spin state of $X$.  If the analog of the factorization 
formula in Eq.~(\ref{A-Z}) is summed over the spin states of the 
$J/\psi$ and $\omega$, it gives
%-----------------
\begin{eqnarray}
\sum_{\rm spins} | {\cal A}_{X\psi \omega} |^2
= \sqrt{8 \pi^2 \Delta_{\psi \omega}/\mu_{\psi \omega}^3} \, 
Z_{\psi \omega},
\end{eqnarray}
%-----------------
where $\Delta_{\psi \omega}$ is the energy gap in Eq.~(\ref{gap-omega}),
$\mu_{\psi \omega}$ is the reduced mass of the $J/\psi$ and $\omega$, 
and $Z_{\psi \omega}$ is the probability for the  $J/\psi \, \omega$
component of $X$.
Squaring both sides of Eq.~(\ref{A-A}) and summing over the spin
states of $J/\psi$ and $\omega$, we get
%-----------------
\begin{eqnarray}
2 m_\omega^2 |G_{X\psi \omega}|^2 = 
16 \pi m_X (m_\psi + m_\omega)  
\sqrt{2 \Delta_{\psi \omega}/\mu_{\psi \omega}} \, 
Z_{\psi \omega}.
\label{G-Z}
\end{eqnarray}
%-----------------
Inserting Swanson's result $Z_{\psi \omega}= 9.6\%$ for 
$E_X = 0.7$ MeV and using the factorization formula in 
Eqs.~(\ref{Gamma:fact}) and (\ref{AXsq}), this reduces to
%-----------------
\begin{eqnarray}
|G_{X\psi \omega}|^2 = 
9.59 \left( {E_X + \Gamma_X^2/(16 E_X) \over 0.7 \, {\rm MeV}} \right)^{1/2}.
\label{Gabs-psiomega}
\end{eqnarray}
%-----------------
Inserting the result into the expression in Eq.~(\ref{Gam-psi3pi}),
we get a quantitative result for the partial width:
%-----------------
\begin{eqnarray}
\Gamma[X\to J/\psi \, \pi^+\pi^-\pi^0]
&=& ( 222 \, {\rm keV} )
\left( {E_X + \Gamma_X^2/(16 E_X) \over 1 \, {\rm MeV}} \right)^{1/2}.
\label{Gam-psi3pi:E} 
\end{eqnarray}
%-----------------

We can use the result in Eq.~(\ref{Gam-psi3pi:E}) to set a lower bound 
on the partial width into $ J/\psi \, \pi^+ \pi^- \pi^0$.  
As a function of the binding energy $E_X$, 
the right side of Eq.~(\ref{Gam-psi3pi:E}) is minimized at 
$E_X = \Gamma_X/4$.  The lower bound on the width is 
$\Gamma_X > 2 \Gamma[D^{*0}] = 136\pm 32$ keV.
Thus the lower bound on the partial width  
into $ J/\psi \, \pi^+ \pi^- \pi^0$ in Swanson's model is about 58 keV.

As is evident in Fig.~\ref{fig:matching}, Swanson did not calculate 
the probability $Z_{\psi \rho}$ for the  $J/\psi \, \rho$
component of $X$ for a binding energy small enough to be in the 
scaling region where $Z_{\psi \rho}$ scales like $E_X^{1/2}$.
If he had, we could use an equation analogous to Eq.~(\ref{G-Z})
to determine $|G_{X\psi \rho}|$.  If we assume that the smallest 
binding energy considered by Swanson is close 
to the scaling region, we can use his value $Z_{\psi \rho}=0.86\%$ 
for $E_X = 0.7$ MeV to estimate $|G_{X\psi \rho}|$.
In the analog of Eq.~(\ref{G-Z}), we should set 
$\Delta_{\psi \rho}=\Delta_{\psi \omega}$ rather than using the 
value $\Delta_{\psi \rho}$ in Eq.~(\ref{gap-rho}), because  
Swanson set $m_\rho = m_\omega$ in his calculation.
The resulting estimate is
%-----------------
\begin{eqnarray}
|G_{X\psi \rho}|^2 \approx
0.86 \left( {E_X + \Gamma_X^2/(16 E_X) \over 0.7 \, {\rm MeV}} \right)^{1/2}.
\label{Gabs-psirho}
\end{eqnarray}
%-----------------
We can insert this estimate into Eq.~(\ref{Gam-psi2pi}) 
to get an estimate of the partial width for decay into
$J/\psi \, \pi^+\pi^-$.
We can also insert this estimate of $|G_{X\psi \rho}|^2$
and the value of $|G_{X\psi \omega}|^2$ from Eq.~(\ref{Gabs-psiomega}) 
into Eqs.~(\ref{Gam-psipigamma}) and (\ref{Gam-psigamma}) 
to get ranges of estimates of the partial widths for the decays into
$J/\psi \, \pi^0 \gamma$ and $J/\psi \, \gamma$.  The ranges
arise from the unknown relative phase between $G_{X\psi \omega}$ 
and $G_{X\psi \rho}$.

%\newpage

\section {Summary}
\label{sec:summary}

Evidence is accumulating that the $X(3872)$ is a loosely-bound 
S-wave molecule corresponding to a
$C=+$ superposition of $D^0\bar D^{*0}$ and $D^{*0}\bar D^{0}$.
Because its binding energy is small compared to the natural 
energy scale associated with pion exchange, this molecule 
has universal properties that are completely determined 
by the large scattering length $a$ in the $C=+$ channel of 
$D^0\bar D^{*0}$ and $D^{*0}\bar D^{0}$.

We have analyzed the decays of $X$ into $J/\psi$ plus light hadrons 
under the assumption that $X$ is a $D D^*$ molecule and that these 
decays proceed through transitions to $J/\psi \, \rho$ and 
$J/\psi \, \omega$.  The differential decay rates were calculated 
in terms of unknown coupling constants $G_{X \psi \rho}$
and $G_{X \psi \omega}$ by using an effective lagrangian that 
reproduces the decays of the light vector mesons.
The dependence on the 
unknown coupling constants enters only through multiplicative 
factors, so the angular distributions are completely determined.

Quantitative predictions of the partial widths for the decays 
of $X$ into $J/\psi$ plus light hadrons require numerical 
values for the  coupling constants $G_{X \psi \rho}$
and $G_{X \psi \omega}$.  We pointed out that the dependence 
of these coupling constants on the binding energy $E_X$ 
and the total width $\Gamma_X$ are determined by factorization 
formulas.  We showed how $|G_{X \psi \omega}|^2$ could be 
determined from the probability $Z_{\psi \omega}$ for the 
$J/\psi \, \omega$ component of $X$ in Swanson's model.  
We used this result to give a quantitative prediction for the 
partial width for $X \to J/\psi \, \pi^+ \pi^- \pi^0$ 
as a function of $E_X$ and $\Gamma_X$.

%\newpage

%%  APPENDIX 
\appendix

\section{Decay amplitudes for vector mesons}
\label{sec:vectormesons}
In this appendix, we present an updated determination 
of the coupling constants in the 
effective lagrangian for the light pseudoscalar 
and vector mesons that was used in Ref.~\cite{Braaten:1989zn}
to calculate the semileptonic branching fractions 
for the $\tau$ lepton.  The same effective lagrangian 
is used in Section~\ref{sec:Xdecay} to calculate the 
decay rates of the $X$ into $J/\psi$ and light hadrons.

The pion decay constant $F_{\pi}= 93$ MeV and the hadron masses 
have all been determined accurately \cite{Eidelman:2004wy}.
The other parameters in the effective lagrangian 
can be determined from the partial widths for decays 
of $\rho^0$, $\rho^\pm$, and $\omega$ given in Table~\ref{tab:data}.
The most useful combinations of the parameters in the amplitudes 
for the decays of the vector mesons into pions in
Eqs.~(\ref{amp-rho}) and (\ref{amp-omega}) are
%-----------------
\begin{subequations}
\begin{eqnarray}
 G_{v \pi \pi} &=& 11.99 \pm 0.06, 
\label{Gvpipi} 
\\
 C_{v 3\pi}+G_{v \pi \pi} C_{vv\pi} F_{\pi}^2/m_v^2 
 &=& (8.03 \pm 0.48)/(16\pi^2),
\label{Cv3pi}
\\
 G_{v \pi \pi} C_{vv\pi} F_{\pi}^2/m_v^2 
 &=& (10.2 \pm 1.3)/(16\pi^2). 
\label{GCvvpi} 
\end{eqnarray} 
\label{coupling1}
\end{subequations}
%-----------------
The coupling constant $G_{v \gamma}$ associated with vector meson dominance 
and the most useful combinations of parameters in the amplitudes for
the radiative decays of the vector mesons in Eqs.~(\ref{amp:pigamma}) are
%-----------------
\begin{subequations}
\begin{eqnarray}
 G_{v \gamma} &=& 14.01\pm 0.11,  
 \label{Gvgam} 
 \\
 C_{v\pi\gamma}+G_{v\gamma}\,C_{vv\pi}\, 
F_\pi^2/m_v^2 &=& (7.99 \pm 0.45)/(16\pi^2), 
\label{Cvpigam} 
\\
G_{v\gamma}C_{vv\pi}\,F_\pi^2/m_v^2 
 &=& (11.9 \pm 1.5)/(16\pi^2). 
 \label{GCvgam}
\end{eqnarray} 
\label{coupling2}
\end{subequations}
%-----------------
The vector meson mixing angle is given by%
\footnote{The cosine of the angle $\theta_v$ here 
is the sine of the vector meson mixing angle 
used in Ref.~\cite{Braaten:1989zn}.}
%-----------------
\begin{equation}
\cos \theta_v = 0.51 \pm 0.01.
\label{costheta}
\end{equation}
%-----------------
Another function of $\theta_v$ that is often encountered is 
$\cos\theta_v+\sqrt{2}\sin\theta_v \approx 1.73\pm 0.01$.
The errors in the parameters in Eqs.~(\ref{coupling1}), 
(\ref{coupling2}), and (\ref{costheta})
are determined using the uncertainties 
in the measurements of the vector meson decay widths only.  
The uncertainties in the
hadron masses and the pion decay constant are negligible in comparison. 
Variations in the parameters associated with 
$U(3)\times U(3)$ symmetry breaking are neglected in this analysis.

%------------------------------------------------------------------------------------
\begin{table}[t]
\begin{tabular}{l|l}
\;Decay mode & \hspace{0.2 cm} Partial width  \\ \hline \hline
$\rho^0\to\pi^+\pi^-$ & \; $ 150.3 \pm 1.6$ MeV  \\ \hline
$\rho^0\to e^+ e^-$  &  \; $ 7.02 \pm 0.11 $ keV  \\ \hline
$\rho^-\to \pi^-\gamma$  &  \; $67.6 \pm 7.5$ keV   \\ \hline
$\omega\to e^+ e^-$  &  \; $ 0.60 \pm 0.02 $ keV  \\ \hline 
$\omega\to \pi^0\gamma$  &  \;  $0.76 \pm 0.05$ MeV  \\ \hline
$\omega\to \pi^0\mu^+\mu^-$  & \; $0.82 \pm 0.20$ keV   \\ \hline
$\omega\to \pi^+\pi^-\pi^0$  & \; $7.56 \pm 0.093$ MeV  \\ 
\hline
\end{tabular}
\caption{
Inputs that are used to determine the coupling constants
in the vector meson decay amplitudes. 
The partial widths are taken from Ref.~\cite{Eidelman:2004wy}.  
\label{tab:data} }
\end{table}
%------------------------------------------------------------------------------------

The inputs that were used to determine the parameters 
in Eqs.~(\ref{coupling1}), (\ref{coupling2}), and (\ref{costheta})
are listed in Table~\ref{tab:data}.
Following Ref.~\cite{Braaten:1989zn}, we determine the parameters 
by the following steps:
\begin{enumerate}

\item  
The coupling constant $G_{v \pi \pi}$ in Eq.~(\ref{Gvpipi}) 
is determined from the partial width for $\rho\to\pi^+\pi^-$: 
%-----------------
\begin{equation}
 \Gamma[\rho\to\pi^+\pi^-]
 =\frac{G_{v \pi \pi}^2 m_\rho}{192\pi} \, 
 \left(1-4 m_\pi^2/m_\rho^2\right)^{3/2}.
\end{equation}
%-----------------

\item
The coupling constant $G_{v \gamma}$ in Eq.~(\ref{Gvgam})
is determined from the partial width for $\rho\to e^+ e^-$:
%-----------------
\begin{equation}
 \Gamma[\rho\to e^+ e^-]
 =\frac{4\pi\alpha_{\rm em}^2\, G_{v \gamma}^2 \,F_{\pi}^4}{3\, m_\rho^3 }.
\end{equation}
%-----------------

\item 
The combination of parameters in Eq.~(\ref{Cvpigam})
is determined from the partial width for $\rho^-\to \pi^-\gamma$:
%-----------------
\begin{equation}
 \Gamma[\rho^-\to\pi^-\gamma]
 =\frac{2\alpha_{\rm em} m_\rho^3}{27 F_{\pi}^2}\, 
\left( C_{v\pi\gamma} + \frac{G_{v\gamma}\,C_{vv\pi}\, 
F_\pi^2}{m_v^2}\right)^2 \, (1-m_\pi^2/m_\rho^2)^3.
\label{Gam:rhopigam}
\end{equation}
%-----------------

\item 
The combination of parameters 
$G_{v\gamma}\,C_{vv\pi}\,F_\pi^2/m_v^2$ 
in Eq.~(\ref{GCvgam}) is determined from the ratio of the partial widths 
for $\omega\to \pi^0\mu^+\mu^-$ and $\omega\to \pi^0\gamma$. 
The possibility of a relative phase between $C_{v\pi\gamma}$ and 
$G_{v\gamma}\,C_{vv\pi}\,F_\pi^2/m_v^2$ 
is ignored.
The partial width for $\omega \to \pi^0\mu^+\mu^-$ is
%-----------------
\begin{equation}
\Gamma[\omega\to \pi^0\mu^+\mu^-] = 
\frac{1}{256\pi^3 m_\omega^3}\int ds_{12} \int ds_{23} 
\mbox{$\overline{\sum}$}|{\cal A}[\omega\to \pi^0\mu^+\mu^-]|^2.
\label{Gam-omegapimu}
\end{equation}
%-----------------
The squared amplitude, averaged over initial spin states and
summed over final spin states, is 
%-----------------
\begin{eqnarray}
\mbox{$\overline{\sum}$} |{\cal A}[\omega\to \pi^0\mu^+\mu^-]|^2
&&=\frac{128\pi^2\alpha_{\rm em}^2 }{9F_\pi^2} \,
(\cos\theta_v+\sqrt{2}\sin\theta_v)^2
\nonumber
\\
&& \hspace{-1cm}
\times \left[(s_{23}^2+4m_\mu^2)
\left((m_\omega^2-s_{23}-m_\pi^2)^2-4m_\pi^2s_{23}\right)
+s_{23}(s_{12}-s_{31})^2\right] 
\nonumber
\\
&& \hspace{-1cm}
\times \frac{1}{s_{23}^2}
\left|C_{v\pi\gamma} 
 +  \frac{ G_{v\gamma}\,C_{vv\pi}\,F_\pi^2 }{m_v^2}
(1-f_\rho( s_{23} ) ) \right|^2,
\end{eqnarray}
%-----------------
where $s_{12}$, $s_{23}$, and $s_{31}$ are the squares of the
invariant masses
of the $\pi^0\mu^+$, $\mu^+\mu^-$, and $\mu^-\pi^0$, respectively.
The partial width for $\omega\to \pi^0\gamma$ is 
%-----------------
\begin{equation}
\Gamma[\omega\to \pi^0\gamma]
=3 (\cos\theta_v + \sqrt{2}\sin\theta_v)^2
{ m_\omega^3 (1-m_\pi^2/m_\omega^2)^3
	\over m_\rho^3 (1-m_\pi^2/m_\rho^2)^3 } \,
\Gamma[\rho^- \to \pi^- \gamma],
\label{Gam-omegapigam}
\end{equation}
%-----------------
where $\Gamma[\rho^- \to \pi^- \gamma]$ is given in
Eq.~(\ref{Gam:rhopigam}).
Note that the factor $(\cos\theta_v + \sqrt{2}\sin\theta_v)^2$
cancels in the ratio of 
Eqs.~(\ref{Gam-omegapimu}) and (\ref{Gam-omegapigam}). 

\item
The combination of parameters 
$G_{v \pi \pi} C_{vv\pi} F_{\pi}^2/m_v^2$ appearing in 
Eq.~(\ref{GCvvpi}) is determined by multiplying the combination 
of parameters in Eq.~(\ref{GCvgam}) by the ratio
$G_{v \pi \pi}/G_{v\gamma}$ obtained from
Eqs.~(\ref{Gvpipi}) and (\ref{Gvgam}).

\item The combination of parameters 
in Eq.~(\ref{Cv3pi}) is determined from the 
ratio of the partial widths for $\omega$ to decay into
$\pi^+\pi^-\pi^0$ and $\pi^0\gamma$ and from the value 
of the combination of parameters in Eq.~(\ref{GCvvpi}).
The possibility of a relative phase between $C_{v 3\pi}$ and
$G_{v \pi \pi} C_{vv\pi}F_{\pi}^2/m_v^2$ is ignored.
The partial width for 
$\omega\to\pi^0\gamma$  is given in Eq.~(\ref{Gam-omegapigam}).
The partial width for $\omega \to \pi^+\pi^-\pi^0$ is
%-----------------
\begin{equation}
\Gamma[\omega\to \pi^0\pi^+\pi^-] = 
\frac{1}{256\pi^3 m_\omega^3}\int ds_{12} \int ds_{23} 
\mbox{$\overline{\sum}$}|{\cal A}[\omega\to \pi^+\pi^-\pi^0]|^2.
\label{Gam-omegapipipi}
\end{equation}
%-----------------
The squared amplitude, averaged over the spin states 
of $\omega$, is 
%-----------------
\begin{eqnarray}
\mbox{$\overline{\sum}$} |{\cal A}[\omega\to \pi^+\pi^-\pi^0]|^2
&&= 
\frac{4 (\cos\theta_v+\sqrt{2}\sin\theta_v)^2}{F_{\pi}^6} \,
\left(s_{12}s_{23}s_{31}-m_{\pi}^2 (m_\omega^2 -m_\pi^2 )^2\right) \,
\nonumber \\
&& \hspace{-2.0cm}
\times
\left|C_{v 3\pi}
+ \frac{ G_{v \pi \pi}C_{vv\pi} F_{\pi}^2 } { m_v^2 } 
\left( 1 -  \mbox{$1\over 3$} \,  
\left[f_\rho(s_{12})+f_\rho(s_{23})+f_\rho(s_{31})\right]
 \right) \right|^2.
\end{eqnarray}
%-----------------
Note that the factor $(\cos\theta_v + \sqrt{2}\sin\theta_v)^2$
cancels in the ratio of 
Eqs.~(\ref{Gam-omegapipipi}) and (\ref{Gam-omegapigam}). 

\item
Finally, the cosine of the vector meson mixing angle in Eq.~(\ref {costheta})
is determined from the ratio of the partial widths for 
$\omega \to e^+ e^-$ and $\rho^0 \to e^+ e^-$:
%-----------------
\begin{equation}
\Gamma[\omega \to e^+ e^-]  =
{\cos^2 \theta_v \, m_\rho^3 \over 3 m_\omega^3} \, 
\Gamma[\rho^0 \to e^+ e^-].
\end{equation}
%-----------------

\end{enumerate}

\begin{acknowledgments}
We thank Mark Wise for useful discussions.
We thank Eric Swanson for providing us with some of his 
numerical results.
This research was supported in part by the Department of Energy 
under grant DE-FG02-91-ER4069.  
\end{acknowledgments}



\begin{thebibliography}{}

%\cite{Choi:2003ue}
\bibitem{Choi:2003ue}
S.~K.~Choi {\it et al.}  [Belle Collaboration],
%``Observation of a new narrow charmonium state in exclusive B+- $\to$
	%K+- pi+
%pi- J/psi decays,''
Phys.\ Rev.\ Lett.\  {\bf 91}, 262001 (2003).
 [arXiv:hep-ex/0309032].

%\cite{Acosta:2003zx}
\bibitem{Acosta:2003zx}
D.~Acosta {\it et al.}  [CDF II Collaboration],
%``Observation of the narrow state X(3872) $\to$ J/psi pi+ pi- in anti-p p
%collisions at s**(1/2) = 1.96-TeV,''
Phys.\ Rev.\ Lett.\  {\bf 93}, 072001 (2004).
 [arXiv:hep-ex/0312021].

%\cite{Abazov:2004kp}
\bibitem{Abazov:2004kp}
V.~M.~Abazov {\it et al.}  [D0 Collaboration],
%``Observation and properties of the X(3872) decaying to J/psi pi+ pi- in p
%anti-p collisions at s**(1/2) = 1.96-TeV,''
Phys.\ Rev.\ Lett.\  {\bf 93}, 162002 (2004)
[arXiv:hep-ex/0405004].

\bibitem{Aubert:2004ns}
  B.~Aubert {\it et al.}  [BABAR Collaboration],
  %``Study of the B $\to$ J/psi K- pi+ pi- decay and measurement of the B $\to$
  %X(3872) K- branching fraction,''
  Phys.\ Rev.\ D {\bf 71}, 071103 (2005)
  [arXiv:hep-ex/0406022].

%\cite{Olsen:2004fp}
\bibitem{Olsen:2004fp}
S.~L.~Olsen  [Belle Collaboration],
%``Search for a charmonium assignment for the X(3872),''
Int.\ J.\ Mod.\ Phys.\ A {\bf 20}, 240 (2005)
[arXiv:hep-ex/0407033].

%\cite{Barnes:2003vb}
\bibitem{Barnes:2003vb}
T.~Barnes and S.~Godfrey,
%``Charmonium options for the X(3872),''
Phys.\ Rev.\ D {\bf 69}, 054008 (2004).
 [arXiv:hep-ph/0311162].

%\cite{Eichten:2004uh}
\bibitem{Eichten:2004uh}
E.~J.~Eichten, K.~Lane and C.~Quigg,
%``Charmonium levels near threshold and the narrow state X(3872) $\to$ pi+ pi-
%J/psi,''
Phys.\ Rev.\ D {\bf 69}, 094019 (2004).
[arXiv:hep-ph/0401210].

%\cite{Quigg:2004nv}
\bibitem{Quigg:2004nv}
C.~Quigg,
%``Quarkonium: New developments,''
arXiv:hep-ph/0403187.

%\cite{Quigg:2004vf}
\bibitem{Quigg:2004vf}
C.~Quigg,
%``The lost tribes of charmonium,''
Nucl.\ Phys.\ Proc.\ Suppl.\  {\bf 142}, 87 (2005)
[arXiv:hep-ph/0407124].

\bibitem{Tornqvist}
%\cite{Tornqvist:2003na}
%\bibitem{Tornqvist:2003na}
N.A.~Tornqvist,
%%``Comment on the narrow charmonium state of Belle 
%%	at 3871.8-MeV as a deuson,''
arXiv:hep-ph/0308277;
%\cite{Tornqvist:2004qy}
%\bibitem{Tornqvist:2004qy}
%N.~A.~Tornqvist,
%``Isospin breaking of the narrow charmonium state of Belle at 3872-MeV as a
%deuson,''
Phys.\ Lett.\ B {\bf 590}, 209 (2004).
%[arXiv:hep-ph/0402237].
%%CITATION = HEP-PH 0402237;%%

%\cite{Voloshin:2003nt}
\bibitem{Voloshin:2003nt}
M.~B.~Voloshin,
 %``Interference and binding effects in decays of possible molecular
	%component
%of X(3872),''
Phys.\ Lett.\ B {\bf 579}, 316 (2004).
 [arXiv:hep-ph/0309307].

%\cite{Wong:2003xk}
\bibitem{Wong:2003xk}
C.~Y.~Wong,
%``Molecular states of heavy quark mesons,''
Phys.\ Rev.\ C {\bf 69}, 055202 (2004).
[arXiv:hep-ph/0311088].
%%CITATION = HEP-PH 0311088;%%
    
%\cite{Braaten:2003he}
\bibitem{Braaten:2003he}
E.~Braaten and M.~Kusunoki,
%``Low-energy universality and the new charmonium resonance at 3870-MeV,''
Phys.\ Rev.\ D {\bf 69}, 074005 (2004).
 [arXiv:hep-ph/0311147].

%\cite{Swanson:2003tb}
\bibitem{Swanson:2003tb}
E.~S.~Swanson,
%``Short range structure in the X(3872),''
Phys.\ Lett.\ B {\bf 588}, 189 (2004).
[arXiv:hep-ph/0311229].
%%CITATION = HEP-PH 0311229;%%

%\cite{Braaten:2004fk}
\bibitem{Braaten:2004fk}
E.~Braaten, M.~Kusunoki and S.~Nussinov,
%``Production of the X(3870) in B meson decay by the coalescence of charm
%mesons,''
Phys.\ Rev.\ Lett.\  {\bf 93}, 162001 (2004).
[arXiv:hep-ph/0404161].

%\cite{Swanson:2004pp}
\bibitem{Swanson:2004pp}
E.~S.~Swanson,
%``Diagnostic decays of the X(3872),''
Phys.\ Lett.\ B {\bf 598}, 197 (2004)
[arXiv:hep-ph/0406080].

%\cite{Voloshin:2004mh}
\bibitem{Voloshin:2004mh}
M.~B.~Voloshin,
%``Heavy quark spin selection rule and the properties of the X(3872),''
Phys.\ Lett.\ B {\bf 604}, 69 (2004)
[arXiv:hep-ph/0408321].

%\cite{Braaten:2004ai}
\bibitem{Braaten:2004ai}
E.~Braaten and M.~Kusunoki,
%``Exclusive production of the X(3872) in B meson decay,''
Phys.\ Rev.\ D {\bf 71}, 074005 (2005)
[arXiv:hep-ph/0412268].

%\cite{Braaten:2005jj}
\bibitem{Braaten:2005jj}
  E.~Braaten and M.~Kusunoki,
  %``Factorization in the production and decay of the X(3872),''
  arXiv:hep-ph/0506087.

%\cite{AlFiky:2005jd}
\bibitem{AlFiky:2005jd}
  M.~T.~AlFiky, F.~Gabbiani and A.~A.~Petrov,
  %``X(3872): Hadronic molecules in effective field theory,''
  arXiv:hep-ph/0506141.
  
  %\cite{Bugg:2004rk}
\bibitem{Bugg:2004rk}
D.~V.~Bugg,
%``Reinterpreting several narrow 'resonances' as threshold cusps,''
Phys.\ Lett.\ B {\bf 598}, 8 (2004).
[arXiv:hep-ph/0406293].

%\cite{Bugg:2004sh}
\bibitem{Bugg:2004sh}
D.~V.~Bugg,
%``The X(3872) and the 3941-MeV peak in omega J/psi,''
Phys.\ Rev.\ D {\bf 71}, 016006 (2005)
[arXiv:hep-ph/0410168].

%\cite{Vijande:2004vt}
\bibitem{Vijande:2004vt}
  J.~Vijande, F.~Fernandez and A.~Valcarce,
  %``Describing non-(q anti-q) candidates,''
  Int.\ J.\ Mod.\ Phys.\ A {\bf 20}, 702 (2005)
  [arXiv:hep-ph/0407136].

%\cite{Close:2003mb}
\bibitem{Close:2003mb}
F.~E.~Close and S.~Godfrey,
%``Charmonium hybrid production in exclusive B meson decays,''
Phys.\ Lett.\ B {\bf 574}, 210 (2003)
[arXiv:hep-ph/0305285].

%\cite{Li:2004st}
\bibitem{Li:2004st}
B.~A.~Li,
%``Is X(3872) a possible candidate of hybrid meson,''
Phys.\ Lett.\ B {\bf 605}, 306 (2005)
[arXiv:hep-ph/0410264].

%\cite{Seth:2004zb}
\bibitem{Seth:2004zb}
K.~K.~Seth,
%``An alternative interpretation of X(3872),''
Phys.\ Lett.\ B {\bf 612}, 1 (2005)
[arXiv:hep-ph/0411122].

%\cite{Maiani:2004vq}
\bibitem{Maiani:2004vq}
L.~Maiani, F.~Piccinini, A.~D.~Polosa and V.~Riquer,
%``Diquark-antidiquarks with hidden or open charm and the nature of X(3872),''
Phys.\ Rev.\ D {\bf 71}, 014028 (2005)
[arXiv:hep-ph/0412098].

%\cite{Close:2003sg}
\bibitem{Close:2003sg}
F.~E.~Close and P.~R.~Page,
%``The D*0 anti-D0 threshold resonance,''
Phys.\ Lett.\ B {\bf 578}, 119 (2004).
 [arXiv:hep-ph/0309253].

%\cite{Pakvasa:2003ea}
\bibitem{Pakvasa:2003ea}
S.~Pakvasa and M.~Suzuki,
%``On the hidden charm state at 3872-MeV,''
Phys.\ Lett.\ B {\bf 579}, 67 (2004).
 [arXiv:hep-ph/0309294].

%\cite{Rosner:2004ac}
\bibitem{Rosner:2004ac}
J.~L.~Rosner,
%``Angular distributions in J/psi (rho0, omega) states near threshold,''
Phys.\ Rev.\ D {\bf 70}, 094023 (2004)
[arXiv:hep-ph/0408334].

%\cite{Kim:2004cz}
\bibitem{Kim:2004cz}
T.~Kim and P.~Ko,
%``Dipion invariant mass spectrum in X(3872) $\to$ J/psi pi pi,''
Phys.\ Rev.\ D {\bf 71}, 034025 (2005)
[arXiv:hep-ph/0405265].

%\cite{Abe:2005iy}
\bibitem{Abe:2005iy}
  K.~Abe,
  %``Experimental constraints on the possible J(PC) quantum numbers of the
  %X(3872),''
  arXiv:hep-ex/0505038.

%\cite{Abe:2005ix}
\bibitem{Abe:2005ix}
  K.~Abe,
  %``Evidence for X(3872) $\to$ gamma J/psi and the sub-threshold decay X(3872)
  %$\to$ omega J/psi,''
  arXiv:hep-ex/0505037.

%\cite{Abe:2003zv}
\bibitem{Abe:2003zv}
K.~Abe {\it et al.}  [Belle Collaboration],
%``Observation of B+ $\to$ psi(3770) K+,''
Phys.\ Rev.\ Lett.\  {\bf 93}, 051803 (2004).
[arXiv:hep-ex/0307061].

%\cite{Abe:2004sd}
\bibitem{Abe:2004sd}
K.~Abe {\it et al.}  [Belle Collaboration],
%``Properties of the X(3872) at Belle,''
arXiv:hep-ex/0408116.

%\cite{Aubert:2004fc}
\bibitem{Aubert:2004fc}
B.~Aubert {\it et al.}  [BABAR Collaboration],
%``Observation of the decay B $\to$ J/psi eta K and search for X(3872)
	%$\to$
%J/psi eta,''
Phys.\ Rev.\ Lett.\  {\bf 93}, 041801 (2004).
[arXiv:hep-ex/0402025].

%\cite{Metreveli:2004px}
\bibitem{Metreveli:2004px}
Z.~Metreveli {\it et al.}  [CLEO Collaboration],
%``Search for X(3872) in untagged gamma gamma fusion and initial state
%radiation production with CLEO III,''
arXiv:hep-ex/0408057.

%\cite{Yuan:2003yz}
\bibitem{Yuan:2003yz}
C.~Z.~Yuan, X.~H.~Mo and P.~Wang,
%``The upper limit of the e+ e- partial width of X(3872),''
Phys.\ Lett.\ B {\bf 579}, 74 (2004).
[arXiv:hep-ph/0310261].

%\cite{Bander:1975fb}
\bibitem{Bander:1975fb}
M.~Bander, G.~L.~Shaw, P.~Thomas and S.~Meshkov,
%``Exotic Mesons And E+ E- Annihilation,''
Phys.\ Rev.\ Lett.\  {\bf 36}, 695 (1976).

%\cite{Voloshin:ap}
\bibitem{Voloshin:ap}
M.~B.~Voloshin and L.~B.~Okun,
%``Hadron Molecules And Charmonium Atom,''
JETP Lett.\  {\bf 23}, 333 (1976).
%[Pisma Zh.\ Eksp.\ Teor.\ Fiz.\  {\bf 23}, 369 (1976)].

%\cite{DeRujula:1976qd}
\bibitem{DeRujula:1976qd}
A.~De Rujula, H.~Georgi and S.~L.~Glashow,
%``Molecular Charmonium: A New Spectroscopy?,''
Phys.\ Rev.\ Lett.\  {\bf 38}, 317 (1977).

%\cite{Nussinov:1976fg}
\bibitem{Nussinov:1976fg}
S.~Nussinov and D.~P.~Sidhu,
%``Loosely Bound States Near The Charm Threshold: Charm Molecules,''
Nuovo Cim.\ A {\bf 44}, 230 (1978).

%\cite{Tornqvist:1993ng}
\bibitem{Tornqvist:1993ng}
N.~A.~Tornqvist,
%``From the deuteron to deusons, an analysis of deuteron - like meson
	%meson
%bound states,''
Z.\ Phys.\ C {\bf 61}, 525 (1994).
[arXiv:hep-ph/9310247].

%\cite{Braaten:2004rn}
\bibitem{Braaten:2004rn}
E.~Braaten and H.~W.~Hammer,
%``Universality in Few-body Systems with Large Scattering Length,''
arXiv:cond-mat/0410417.

%\cite{Tornqvist:1991ks}
%\bibitem{Tornqvist:1991ks}
%N.~A.~Tornqvist,
%``Possible large deuteron - like meson meson states bound by pions,''
%Phys.\ Rev.\ Lett.\  {\bf 67}, 556 (1991).

%\cite{Braaten:2004rw}
%\bibitem{Braaten:2004rw}
%  E.~Braaten and M.~Kusunoki,
%  %``Production of the X(3870) at the Upsilon(4S) by the coalescence of  charm
%  %mesons from B decays,''
%  Phys.\ Rev.\ D {\bf 69}, 114012 (2004)
%  [arXiv:hep-ph/0402177].
  
 %\cite{Eidelman:2004wy}
\bibitem{Eidelman:2004wy}
S.~Eidelman {\it et al.}  [Particle Data Group Collaboration],
%``Review of particle physics,''
Phys.\ Lett.\ B {\bf 592}, 1 (2004).

%\cite{Braaten:1989zn}
\bibitem{Braaten:1989zn}
E.~Braaten, R.~J.~Oakes and S.~M.~Tse,
%``An Effective Lagrangian Calculation Of The Semileptonic Decay Modes Of The
%Tau Lepton,''
Int.\ J.\ Mod.\ Phys.\ A {\bf 5}, 2737 (1990);
%%CITATION = IMPAE,A5,2737;%%
%\cite{Braaten:1987jh}
%\bibitem{Braaten:1987jh}
%E.~Braaten, R.~J.~Oakes and S.~M.~Tse,
%``An Effective Lagrangian Calculation Of The Pi Pi Eta Decay Mode Of The Tau
%Lepton,''
Phys.\ Rev.\ D {\bf 36}, 2188 (1987).
%%CITATION = PHRVA,D36,2188;%%

%\cite{Cohen:2004kf}
\bibitem{Cohen:2004kf}
  T.~D.~Cohen, B.~A.~Gelman and U.~van Kolck,
  %``An effective field theory for coupled-channel scattering,''
  Phys.\ Lett.\ B {\bf 588}, 57 (2004).
  [arXiv:nucl-th/0402054].

%\cite{Savage:1996tb}
\bibitem{Savage:1996tb}
  M.~J.~Savage,
  %``The Delta Delta intermediate state in (1)S(0) N N scattering  from
  %effective field theory,''
  Phys.\ Rev.\ C {\bf 55}, 2185 (1997)
  [arXiv:nucl-th/9611022].

\end{thebibliography}
\end{document}